\documentclass[pdflatex,sn-nature]{sn-jnl}
\usepackage{graphicx}%
\usepackage{multirow}%
\usepackage{amsmath,amssymb,amsfonts}%
\usepackage{amsthm}%
\usepackage{mathrsfs}%
\usepackage[title]{appendix}%
\usepackage{xcolor}%
\definecolor{mygray}{gray}{0.9}
\usepackage{textcomp}%
\usepackage{manyfoot}%
\usepackage{booktabs}%
\usepackage{algorithm}%
\usepackage{algorithmicx}%
\usepackage{algpseudocode}%
\usepackage{listings}%
\usepackage{soul}  
\usepackage{bm}
\usepackage{upgreek}

\definecolor{mygray}{gray}{0.9}
\begin{document}
\title[Article Title]{Bound states in the continuum in multilayered time-varying metasurfaces}

\author{Puneet Garg$^\textrm{1*}$}
\author{Michael Plum$^\textrm{2}$}
\author{Carsten Rockstuhl$^\textrm{1,3,4}$}

\affil{\centering{\textsuperscript{1}Institute of Theoretical Solid State Physics, Karlsruhe Institute of Technology, Kaiserstrasse 12, Karlsruhe, 76131, Germany\\
\textsuperscript{2}Institute for Analysis, Karlsruhe Institute of Technology, Kaiserstrasse 12, Karlsruhe, 76131, Germany\\
\textsuperscript{3}Institute of Nanotechnology, Karlsruhe Institute of Technology, Kaiserstrasse 12, Karlsruhe, 76131, Germany\\

\textsuperscript{4}Center for Integrated Quantum Science and Technology (IQST), Karlsruhe Institute of Technology, Kaiserstrasse 12, Karlsruhe, 76131, Germany\\

\vspace{1em}

\noindent\textbf{*Corresponding author E-mail:} 
\href{mailto:puneet.garg@kit.edu}{puneet.garg@kit.edu}}}

\abstract{
Time-varying metamaterials involve a rapid temporal modulation of the permittivity, often at frequencies comparable to the oscillation frequency of light. However, pronounced physical effects at low modulation amplitudes are observed only when resonances sustained in the metamaterials are utilized. This requires an additional spatial structuring. Here, we demonstrate the first exploitation of bound states in the continuum (BICs) in such spatio-temporal metamaterials consisting of a multilayered metasurface. Leveraging Fabry-Perot BICs in a metasurface-based cavity, we realize polarization-insensitive scattering anomalies such as exceptional points (EPs), coherent perfect absorption (CPA), and lasing at extremely small modulation amplitudes. In a second example, by utilizing symmetry-protected BICs and breaking time-reversal symmetry of a multilayered metasurface, we obtain strong nonreciprocal behavior. Harnessing nonreciprocity, we further demonstrate a device capable of one-way monochromatic light transmission at perturbative modulation amplitudes. Our contribution establishes BIC-enabled spatio‑temporal metamaterials as a scalable platform for low-power, tunable light–matter interactions, opening new pathways toward practical nonreciprocal photonic devices, dynamic wave control, and on-chip optical signal processing.}




\maketitle

\noindent Time-varying metamaterials are dynamic media that are formed by rapidly modulating the permittivity in time, often at frequencies comparable with the oscillation frequency of light \cite{galiffi2022photonics,Lustig2023photonic}. While the history of time-varying metamaterials dates back to the $1950$s \cite{morgenthaler1958velocity}, in recent decades, there has been a resurgence of interest in exploring their properties \cite{galiffi2022photonics}. A plethora of exciting physical phenomena have been reported, including time reflection \cite{Moussa2023observation}, time refraction \cite{lustig2023time}, parametric amplification \cite{zurita-sanchez2009reflection,lyubarov2022amplified}, magnetless nonreciprocity \cite{shaltout2015time,asadchy2020tutorial}, frequency conversion \cite{amra2024linear}, subluminal Cherenkov radiation \cite{dikopoltsev2022light}, and the emergence of momentum bandgaps \cite{wang2023metasurface,reyes-ayona2015observation,Li2023stationary}. Complementing theoretical advances, experimental platforms involving transparent conducting oxides (TCOs) and dielectrics have been promising in demonstrating several of these phenomena \cite{harwood2025space,galiffi2026optical,shilkin2024ultrafast,liberal2017near}. 

Early works on time-varying metamaterials largely investigated spatially homogeneous media, yielding important fundamental insights \cite{galiffi2022photonics}. However, recent studies have increasingly focused on combining the effects of spatial structuring and temporal modulation \cite{Sharabi2022spatiotemporal, Stefanou2021light, ptitcyn2023floquet,liberal2026synthetic}. In that realm, spheres \cite{Stefanou2021light, ptitcyn2023floquet,asadchy2022parametric,sadafi2023dynamic,verde2026optical}, cylinders \cite{Stefanou2023light}, slabs \cite{valero2026revealing,zurita-sanchez2009reflection,globosits2024photonic,rizza2024harnessing}, and metasurfaces \cite{garg2022modeling,garg2025photonic,panagiotidis2023optical,wang2025expanding,wang2023metasurface,Paz2026lattice} made from time-varying media are continually being explored. Spatial structuring provides a pathway to exploit nanophotonic resonances in time-varying metamaterials \cite{asadchy2022parametric,garg2025photonic,wang2025expanding}. It is the resonant enhancement of light–matter interactions in such metamaterials that enables the observation of strong physical effects even with perturbative modulation amplitudes. Essentially, by trapping light within the nanostructures, these resonances prolong the interaction of light with time-varying media, thereby intensifying the resulting effects. Nanophotonic resonances have already been utilized for expanding momentum bandgaps \cite{garg2025photonic,wang2025expanding}, enhancing parametric amplification \cite{asadchy2022parametric,valero2026revealing}, and enhancing time reflection \cite{rawat2026generation}. 

A class of resonances particularly at the forefront of nanophotonics is that of bound states in the continuum (BICs) \cite{hsu2016bound, koshelev2018asymmetric}. Such resonances are interesting because, in the limit of vanishing material losses, they exhibit diverging radiative quality factors (Q-factors). Hence, ideally, they can indefinitely intensify the light-matter interaction. BICs have revolutionized many areas of photonics, such as ultrasensitive sensing \cite{thapa2025leveraging}, low-threshold nanolasing \cite{ren2022low-threshold}, and nonlinear light generation \cite{Vabishchevich2023nonlinear}, among many others \cite{kang2023applications}. However, the exploitation of BICs in time-varying structures has remained largely unexplored \cite{garg2025photonic,hayran2021capturing}. In fact, to the best of our knowledge, BICs in multilayered time-varying metasurfaces that essentially form metamaterials have never been considered.

In our contribution, we apply the physics of BICs to multilayered time-varying metasurfaces. To demonstrate the generality of our work, we consider Fabry-Perot BICs (FP-BICs) as well as symmetry-protected BICs (SP-BICs) \cite{koshelev2023bound}. Using the example of FP-BICs in a metasurface-based cavity, we show that scattering anomalies, such as exceptional points (EPs), coherent perfect absorption (CPA), and lasing, can be achieved with a \textit{polarization-insensitive} operation. We find that these scattering anomalies emerge at extremely small modulation amplitudes. In a second example involving SP-BICs, we show that breaking the time-reversal symmetry of multilayered metasurfaces can engineer a strong nonreciprocal response. Building on this nonreciprocity, we design a device capable of one-way light transmission. The proposed nonreciprocal device requires extremely low modulation amplitudes. Furthermore, despite time modulation, a careful choice of system parameters ensures that the nonreciprocal transmission remains monochromatic. Such a monochromatic behavior is in stark contrast to other time modulation-based nonreciprocal devices reported in the literature \cite{garg2025inverse,sedeh2022optical,panagiotidis2023optical,ramaccia2020electromagnetic}.  

We emphasize that although the examples discussed in this work operate in the perturbative regime of temporal modulation, the underlying computational framework is general and remains valid even for large modulation amplitudes.

\section*{Results}
The core of our contribution is organized into three sections. We begin by presenting the theoretical framework for describing spatio-temporal metamaterials composed of stacked layers of time-varying metasurfaces. We then discuss two applications: the exploitation of scattering anomalies associated with FP-BICs, and nonreciprocal transmission enabled by SP-BICs.

\subsection*{Theory}
\begin{figure*}
\centerline{\includegraphics[width= 1\columnwidth,trim=0.1 0.1 0.1 0.1,clip]{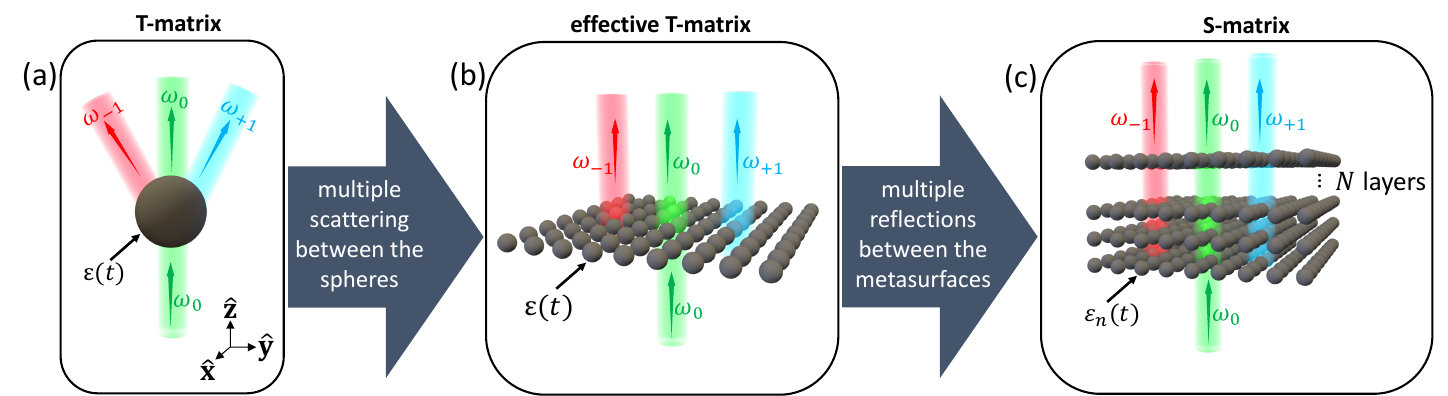}}
\caption{\textbf{Time-varying scattering structures}. (a) A time-varying sphere made from a material with permittivity $\varepsilon(t)=1+\chi_\mathrm{st}[1+M\mathrm{cos}(\Omega t+\phi)]$. The sphere is numerically modeled using the T-matrix. (b) A metasurface made from a periodic arrangement of time-varying spheres in 2D. The scattering response of the metasurface is modeled using the effective T-matrix by considering the multiple scattering between the spheres. (c) A stack of $N$ time-varying metasurfaces, where each sphere in the $n^\mathrm{th}$ metasurface has the permittivity $\varepsilon_n(t)=1+\chi_\mathrm{st}[1+M_n\mathrm{cos}(\Omega t+\phi_n)]$. The scattering response of the multilayered metasurface is modeled using the S-matrix by considering the multiple reflections between the metasurfaces.}
\label{fig:concept}
\end{figure*}

In this section, we revisit the theory for computing the scattering properties of various geometries shown in Fig.~\ref{fig:concept} \cite{ptitcyn2023floquet,garg2022modeling,Stefanou2021light}. Specifically, we discuss the transition matrix (T-matrix) and scattering matrix (S-matrix) methods \cite{waterman1965proceedings,Krasnok2019anomalies}. As it is necessary for time-varying media, besides the expressions for positive frequencies, we also provide appropriate symmetry relations that are needed to evaluate these matrices at negative frequencies. We begin with an isolated time-varying sphere (see Fig.~\ref{fig:concept}(a)). We use the T-matrix method to analytically model the sphere. This method has been introduced in \cite{ptitcyn2023floquet,Stefanou2021light}. The sphere is made from a material with permittivity $\varepsilon(t)=1+\chi_\mathrm{st}[1+M\mathrm{cos}(\Omega t+\phi)]$. Here, $\chi_\mathrm{st}$ is the susceptibility of the static medium, $M$ is the modulation amplitude, $\Omega$ is the modulation frequency, and $\phi$ is the modulation phase. In general, time modulation couples fields with frequencies on a comb defined by $\omega_j=\omega+j\Omega$, where $j$ is an integer with $j\in[-J,J-1]$. Furthermore, $\omega$ is the Floquet frequency, and $2J$ is the total number of frequencies in the comb. Note that in the static case, i.e., $M=0$ and $\Omega=0$, we retain only the central frequency (with $j=0$) in the comb. 

To solve the scattering problem, we expand the incident field ${\mathbf{E}}^\mathrm{inc}(\mathbf{r},t)$ and the scattered field $\mathbf{E}^{\mathrm{sca}}(\mathbf{r}, t)$ for the time-varying sphere in a basis of vector spherical waves (VSWs) as
\begin{subequations}\label{eq:inc_sca}
\begin{align}
    {\mathbf{E}}^{\mathrm{inc}}(\mathbf{r},t)&=\sum_{jlm s}{A}_{jlm s}^{\mathrm{inc}}\mathbf{F}^{(1)}_{lms}(k_j\mathbf{r})\mathrm{e}^{-i\omega_j t}\,,\label{eq:E_inc}\\
    {\mathbf{E}}^{\mathrm{sca}}(\mathbf{r},t)&=\sum_{jlm s}{A}_{jlm s}^{\mathrm{sca}}\mathbf{F}^{(3)}_{lm s}(k_j\mathbf{r})\mathrm{e}^{-i\omega_jt}\,,\label{eq:E_sca}
    \end{align}
    \end{subequations}
    \noindent
where $k_j=\omega_j/c_0$, with $c_0$ being the speed of light in a vacuum. Note that for $\omega_j<0$, $k_j<0$. Furthermore, ${A}_{jlm s}^{\mathrm{inc}} \textrm{ and } {A}_{jlm s}^{\mathrm{sca}}$ represent the incident and scattered field coefficients, respectively. Moreover, $\mathbf{F}^{(1)}_{lm s}(k_j\mathbf{r})$  and $\mathbf{F}^{(3)}_{lm s}(k_j\mathbf{r})$ are the regular and radiating VSWs, respectively, with total angular momentum $l=1,2,3...,l_\mathrm{max}$, $z$-component of angular momentum $m= -l, -l+1,...,l$, and polarization $s= 0,1$. Here, $s=0$ represents the transverse-electric (TE), and $s=1$ represents the transverse-magnetic (TM) polarization, respectively. The parameter $l_\mathrm{max}$ expresses the maximum multipolar order retained in the expansion.

The coefficient vectors $\mathbf{A}^\mathrm{inc}$ and $\mathbf{A}^\mathrm{sca}$ are linked by the T-matrix $\mathbf{T}_0(\omega)$ as \cite{ptitcyn2023floquet}

\begin{eqnarray}
\mathbf{A}^\mathrm{sca} = \mathbf{T}_0(\omega)\,\mathbf{A}^\mathrm{inc}\, .\label{eq:Tmat_sph}
\end{eqnarray}
Here, $\mathbf{T}_0(\omega)$ is a square polychromatic matrix with dimension $4J l_\mathrm{max}(l_\mathrm{max}+2)$. 

Next, we compute the effective T-matrix of a metasurface made from a two-dimensional (2D) periodic arrangement of time-varying spheres (see Fig.~\ref{fig:concept}(b)). Such a computation requires a rigorous treatment of the lattice sum to account for all multiple scattering contributions within the periodic structure \cite{ewald1921die,beutel2021efficient}. The resulting T-matrix is given as \cite{beutel2021efficient,beutel2024treams,garg2022modeling}

\begin{equation}
\mathbf{T}_\mathrm{eff}(\omega,\mathbf{k}_\parallel)= \left(\mathbf{I}-{\mathbf{T}_{0}(\omega)}\sum_{\mathbf{R}\neq0}\mathbf{C}(\omega,\mathbf{R})\hspace{1pt}\mathrm{e}^{i\mathbf{k_{{\parallel}}}\cdot\mathbf{R}}\right)^{-1}{\mathbf{T}_{0}(\omega)}\,,\label{eq:Teff}
\end{equation}
where $\mathbf{I}$ is the identity matrix, $\mathbf{k}_\parallel$ is the in-plane wavevector, $\mathbf{R}$ is a lattice vector, and $\mathbf{C}(\omega,\mathbf{R})$ is the matrix that contains the translation coefficients of VSWs \cite{beutel2021efficient, cruzan1962translational,stein1961addition}. Here, in general, the determinant of the matrix inside the parentheses is non-zero. A vanishing determinant defines the condition for obtaining the eigenmodes of the metasurface. Furthermore, the matrix $\mathbf{C}(\omega,\mathbf{R})$ has a block diagonal form with the $j^\mathrm{th}$ block containing the translation coefficients for the frequency $\omega_j$. Specifically, 
\begin{equation}
\mathbf{C}(\omega,\mathbf{R}) = 
\begin{bmatrix}
\mathbf{c}(-k_{-J}\mathbf{R}) & 0      & \cdots & 0 \\
0      & \mathbf{c}(-k_{-J+1}\mathbf{R}) & \cdots & 0 \\
\vdots & \vdots & \ddots & \vdots \\
0      & 0      & \cdots & \mathbf{c}(-k_{J-1}\mathbf{R})
\end{bmatrix}\,.
\end{equation}
Here, the elements of $\mathbf{c}(-k_j\mathbf{R})$ satisfy the symmetry relation \cite{Kim2004symmetry} (see the Supplementary Sections~\ref{supp-sec:VSW_symmetry}--\ref{supp-sec:C_symmetry} for a detailed derivation)

\begin{equation}
\label{eq:cmatsymm}    
    c^{\{lms\},\{l'm's'\}}(-k_j\mathbf{R})=(-1)^{l-l'+m'-m+\delta_{s 1}-\delta_{s' 1}}\left[{c}^{\{l(-m)s\},\{l'(-m')s'\}}(-(-k^*_j)\mathbf{R})\right]^*\,,
\end{equation}
where $\delta$ is the Kronecker delta distribution. For $k_j\in\mathbb{R}$, Eq.~\eqref{eq:cmatsymm} enables the computation of $\mathbf{c}(-k_j\mathbf{R})$ for $k_j<0$ from the corresponding expression at $k_j>0$. In combination with this symmetry relation, the matrix $\mathbf{C}(\omega,\mathbf{R})$ and the associated lattice sum in Eq.~\eqref{eq:Teff} can be directly computed using open-source codes written for static metamaterials, such as \colorbox{mygray}{\texttt{treams}} \cite{beutel2024treams}. 

It is convenient to transform the effective T-matrix into the S-matrix \cite{globosits2025exceptional,garg2022modeling}. The S-matrix of a metasurface is usually expressed in the plane wave (PW) basis. Therefore, we expand the incident field ${\mathbf{E}}^\mathrm{inc}(\mathbf{r}, t)$ and outgoing field $\mathbf{E}^{\mathrm{out}}(\mathbf{r}, t)$ for the time-varying metasurface in a basis of PWs as \cite{beutel2021efficient,garg2022modeling}

\begin{subequations}\label{eq:Smat_fields}
\begin{align}
\mathbf{E}^\mathrm{inc}(\mathbf{r},t)
&=
\sum_{\substack{j\mathbf{g}\alpha d\\ \textrm{s.t. } \omega_j>0}}
U^\mathrm{inc}_{j\mathbf{g}\alpha d}
e^{i(\mathbf{k}_{j\mathbf{g}\alpha d}\cdot\mathbf{r}-\omega_j t)}
\,\hat{\mathbf{e}}_{\alpha}(\mathbf{k}_{j\mathbf{g}\alpha d})+
\sum_{\substack{j\mathbf{g}\alpha d\\ \textrm{s.t. } \omega_j<0}}
U^\mathrm{inc}_{j\mathbf{g}\alpha d}
e^{-i(\mathbf{k}^*_{j\mathbf{g}\alpha d}\cdot\mathbf{r}+\omega_j t)}
\,\hat{\mathbf{e}}^*_{\alpha}(\mathbf{k}_{j\mathbf{g}\alpha d})\,,
\label{eq:E_in}
\\[1em]
\mathbf{E}^\mathrm{out}(\mathbf{r},t)
&=
\sum_{\substack{j\mathbf{g}\alpha d\\ \textrm{s.t. }\omega_j>0}}
U^\mathrm{out}_{j\mathbf{g}\alpha d}
e^{i(\mathbf{k}_{j\mathbf{g}\alpha d}\cdot\mathbf{r}-\omega_j t)}
\,\hat{\mathbf{e}}_{\alpha}(\mathbf{k}_{j\mathbf{g}\alpha d})+
\sum_{\substack{j\mathbf{g}\alpha d\\ \textrm{s.t. }\omega_j<0}}
U^\mathrm{out}_{j\mathbf{g}\alpha d}
e^{-i(\mathbf{k}^*_{j\mathbf{g}\alpha d}\cdot\mathbf{r}+\omega_j t)}
\,\hat{\mathbf{e}}^*_{\alpha}(\mathbf{k}_{j\mathbf{g}\alpha d})\,,
\label{eq:E_out}
\\[1em]
\intertext{where}
\mathbf{k}_{j\mathbf{g}\alpha d}
&=
\begin{cases}
\left(\mathbf{k}_\parallel+\mathbf{g}\right)
+\hat{\mathbf{z}}\,(-1)^{\delta_{d\downarrow}}\,\Gamma_z,
\quad
\text{with }
\Gamma_z=\sqrt{k_j^2-\left(\mathbf{k}_\parallel+\mathbf{g}\right)^2},
\quad
\text{if }\omega_j>0,
\\[1em]
\left(-\mathbf{k}_\parallel+\mathbf{g}\right)
+\hat{\mathbf{z}}\,(-1)^{\delta_{d\downarrow}}\,\Gamma_z,
\quad
\text{with }
\Gamma_z=\sqrt{(-k_j)^2-\left(-\mathbf{k}_\parallel+\mathbf{g}\right)^2},
\quad
\text{if }\omega_j<0.
\end{cases}
\end{align}
\end{subequations}

\noindent
The variable $d=\uparrow,\downarrow$ for the upward and downward propagating fields, respectively, and $\alpha=0,1$ for the TE and TM-polarized PWs, respectively. Furthermore, $\mathbf{k}_{j\mathbf{g}\alpha d}$ is a wavevector, $\mathbf{g}$ is a vector characterizing the diffraction order, and $\hat{\mathbf{e}}_\alpha (\mathbf{k}_{j\mathbf{g}\alpha d})$ is a unit vector (see the Supplementary Section~\ref{supp-sec:unit_vectors}). Moreover, $U^\mathrm{inc}_{j\mathbf{g}\alpha d}$ and $U^\mathrm{out}_{j\mathbf{g}\alpha d}$ are the incident and outgoing field coefficients in the PW basis, respectively. We retain $G$ diffraction orders in the above expansion. Here, the branch cuts between $\omega_j>0$ and $\omega_j<0$ arise due to the specific form of the scattered field from the metasurface in the PW basis for an incident unit PW (see Supplementary Section~\ref{supp-sec:scattered_PW}).

The coefficients $\mathbf{U}^\mathrm{inc}$ and $\mathbf{U}^\mathrm{out}$ are linked by the S-matrix $\mathbf{S}(\omega,\mathbf{k}_\parallel)$ as \cite{garg2022modeling, panagiotidis2023optical}

\begin{equation}
\label{eq:Smatrix}
\begin{pmatrix}
\mathbf{U}^\mathrm{out}_{\downarrow}\\
\mathbf{U}^\mathrm{out}_{\uparrow}
\end{pmatrix}
=
\underbrace{
\begin{pmatrix}
\mathbf{S}_{\downarrow\uparrow} & \mathbf{S}_{\downarrow\downarrow}\\
\mathbf{S}_{\uparrow\uparrow} & \mathbf{S}_{\uparrow\downarrow}
\end{pmatrix}
}_{\mathbf{S}(\omega,\mathbf{k}_\parallel)}
\begin{pmatrix}
\mathbf{U}^\mathrm{inc}_{\uparrow}\\
\mathbf{U}^\mathrm{inc}_{\downarrow}
\end{pmatrix}.
\end{equation}
Here, $\mathbf{S}(\omega,\mathbf{k}_\parallel)$ is a square matrix with dimension $8J G$, and $\mathbf{S}_{dd'}$ denotes the sub-block mapping incident coefficients $\mathbf{U}^{\mathrm{inc}}_{d'}$ to outgoing components $\mathbf{U}^{\mathrm{out}}_{d}$. The specific arrangement of the sub-blocks in $\mathbf{S}(\omega,\mathbf{k}_\parallel)$ is chosen according to \cite{novitsky2020unambiguous}. We discuss the detailed steps to compute the S-matrix of a time-varying metasurface from its effective T-matrix in the Supplementary Sections~\ref{supp-sec:VSW_symmetry}--\ref{supp-sec:scattered_PW}. 

Finally, to compute the S-matrix of a composite system formed by stacking the metasurface layers, multiple reflections between the layers have to be considered (see Fig.~\ref{fig:concept}(c)). We use the S-matrix star product to account for such reflections \cite{Redheffer1959inequalities,Rumpf2011improved,stefanou2000MULTEM, stefanou1998heterostructures} (see the Supplementary Section~\ref{supp-sec:star_product}). The total S-matrix of a stack of $N$ metasurfaces is given by 

\begin{align}
    \label{eq:star-product}\mathbf{S}=\mathbf{S}^{(0)}\star\mathbf{S}_\mathrm{prop,\mathbf{r}_0}\star\mathbf{S}^{(1)}\star\mathbf{S}_\mathrm{prop,\mathbf{r}_1}\star\mathbf{S}^{(2)}...\star\mathbf{S}_\mathrm{prop,\mathbf{r}_{N-2}}\star\mathbf{S}^{(N-1)}.
\end{align}
Here, $\mathbf{S}^{(i)}$ denotes the S-matrix of the $i^\textrm{th}$ metasurface, and $\mathbf{S}_\mathrm{prop,\mathbf{r}_i}$ takes into account the free space propagation between $i^\textrm{th}$ and $(i+1)^\textrm{th}$ metasurface. Note that here and in the following, we omit the argument $(\omega,\mathbf{k}_\parallel)$ of the S-matrices for brevity. Assuming that  $(i+1)^\textrm{th}$ metasurface is spatially located above the $i^\textrm{th}$ metasurface, the elements of $\mathbf{S}_\mathrm{prop,\mathbf{r}_i}$ are given by
\begin{equation}
S_{\mathrm{prop},{\mathbf{r}_i}}^{\{j\mathbf{g}\alpha d\},\{j'\mathbf{g}'\alpha'd'\}} =
\begin{cases}
\delta_{jj'}\delta_{dd'}\delta_{\alpha\alpha'}\delta_{\mathbf{g}\mathbf{g}'}e^{i(-1)^{\delta_{d\downarrow}}\mathbf{k}_{j\mathbf{g}\alpha d}\cdot\mathbf{r}_i}\,,\textrm{if }\omega_j>0\,,
\\[1em]
\displaystyle
\delta_{jj'}\delta_{dd'}\delta_{\alpha\alpha'}\delta_{\mathbf{g}\mathbf{g}'}e^{-i(-1)^{\delta_{d\downarrow}}\mathbf{k}^*_{j\mathbf{g}\alpha d}\cdot\mathbf{r}_i}\,,\textrm{if }\omega_j<0\,,
\end{cases}
\end{equation}
where $\mathbf{r}_i$ is the spatial displacement vector drawn from the $i^\textrm{th}$ to $(i+1)^\textrm{th}$ metasurface.

We can also transform the S-matrix into a photon number flux basis. The corresponding transformation is given by \cite{globosits2025exceptional}

\begin{equation}
\tilde{S}_\mathrm{pp}^{\{j\mathbf{g}\alpha d\},\{j'\mathbf{g}'\alpha'd'\}} =
\left| \frac{\omega_{j'}}{\omega_j} \right|
\sqrt{ \left|\frac{k_{z,j\mathbf{g}\alpha d}}{k_{z,j'\mathbf{g}'\alpha' d'}} \right|}\,{S}_\mathrm{pp}^{\{j\mathbf{g}\alpha d\},\{j'\mathbf{g}'\alpha'd'\}}\,. 
\end{equation}
Here, $\tilde{\mathbf{S}}_\mathrm{pp}$ and ${\mathbf{S}}_\mathrm{pp}$ are reduced matrices that only contain the scattering information about the propagating waves. As such, the entries corresponding to incident and outgoing evanescent waves are not retained. Physically, these reduced S-matrices correspond to the case where the source and the detector are placed in the far-field of the metasurface. Note that, for a composite system, while accounting for the multiple reflections between metasurfaces, the full S-matrices $\mathbf{S}^{(i)}$ should be used in Eq.~\eqref{eq:star-product}. Only \textit{after} such a computation, the final S-matrix is reduced to ${\mathbf{S}}_\mathrm{pp}$ or $\tilde{\mathbf{S}}_\mathrm{pp}$. In particular, the reduced matrix $\tilde{\mathbf{S}}_\mathrm{pp}$ is convenient as it satisfies the pseudounitarity condition \cite{globosits2024photonic,globosits2025exceptional} 
\begin{equation}\label{eq:pseudounitarity_S}
\tilde{\mathbf{S}}_\mathrm{pp}^\dagger{\mathbf{V}}\tilde{\mathbf{S}}_\mathrm{pp}={\mathbf{V}}\,, 
\end{equation}
where $\dagger$ denotes the conjugate transpose. Additionally, for a reciprocal configuration, it also satisfies the reciprocity condition \cite{globosits2024photonic,globosits2025exceptional} 
\begin{equation}\label{eq:reciprocity_S}
{\mathbf{V}}\tilde{\mathbf{S}}_\mathrm{pp}^\mathrm{T}{\mathbf{V}}=\tilde{\mathbf{S}}_\mathrm{pp}\,. 
\end{equation}
Here, ${\mathbf{V}}$ is a diagonal matrix with entries $\pm1$ depending on the sign of $\omega_j$, i.e., ${V}^{\{j\mathbf{g}\alpha d\},\{j'\mathbf{g}'\alpha'd'\}} =
\frac{\omega_{j}}{|\omega_j|}
\delta_{jj'}\delta_{\mathbf{g}\mathbf{g}'}\delta_{\alpha\alpha'}\delta_{dd'}\,$ \cite{globosits2024photonic,globosits2025exceptional}.
Note that combining Eq.~\eqref{eq:pseudounitarity_S} and Eq.~\eqref{eq:reciprocity_S} yields $\tilde{\mathbf{S}}_\mathrm{pp}^\dagger~=~\left(\tilde{\mathbf{S}}_\mathrm{pp}^{-1}\right)^\mathrm{T}$.
\subsection*{Polarization-insensitive scattering anomalies based on Fabry-Perot BICs}
\begin{figure*}
\centerline{\includegraphics[width= 1\columnwidth,trim=0.1 0.1 0.1 0.1,clip]{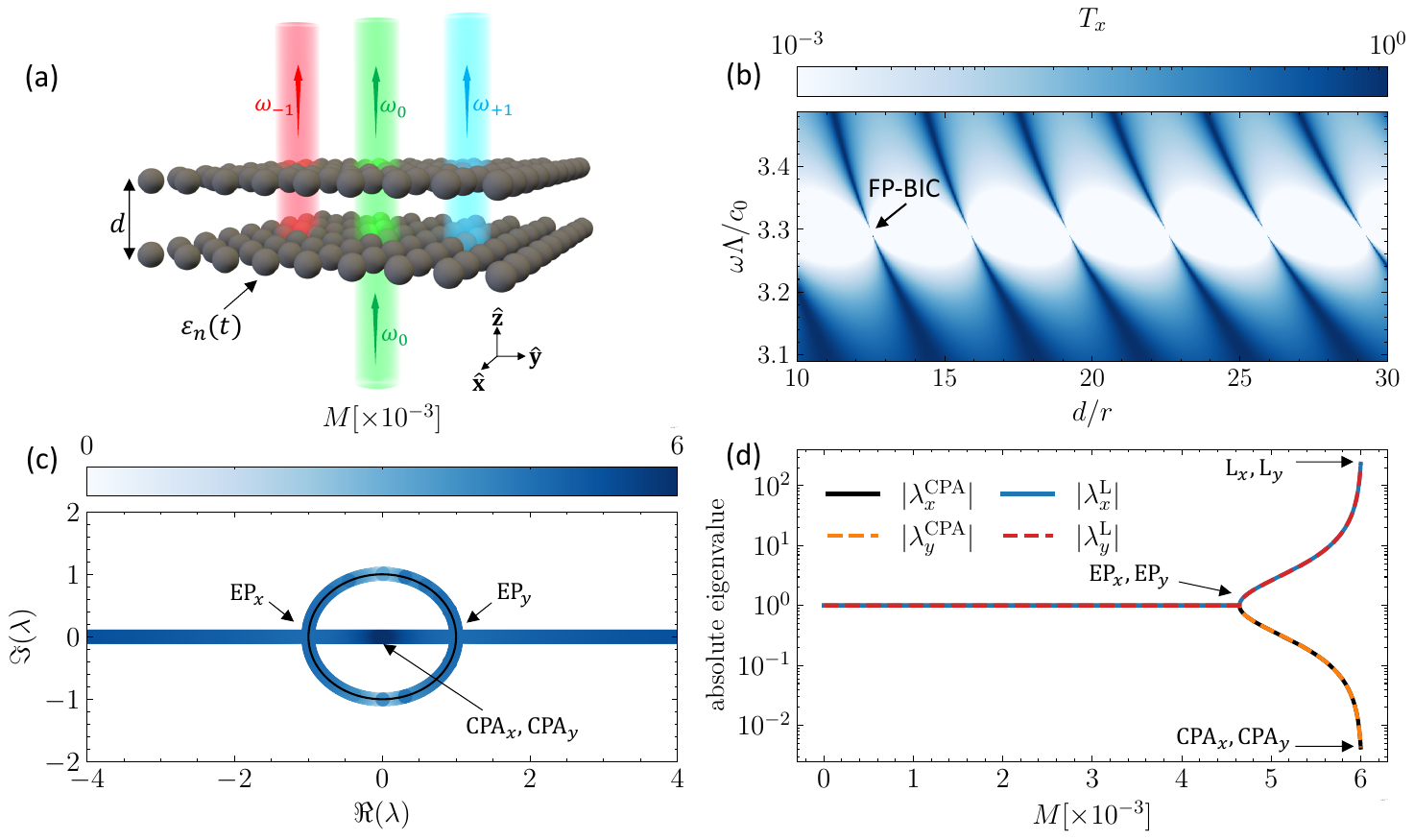}}
\caption{\textbf{Polarization-insensitive exceptional points, coherent perfect absorption (CPA), and lasing in a Fabry-Perot cavity}. (a) A Fabry-Perot cavity formed by stacking two identical time-varying metasurfaces. The permittivity of each layer is $\varepsilon(t)=1+\chi_\mathrm{st}[1+M\mathrm{cos}(\Omega t)]$. (b) The transmissivity $T_x$ of the FP cavity as a function of frequency $\omega$ and the inter-layer distance $d$ under static conditions. The cavity hosts a series of FP-BICs. (c) The real and imaginary parts of the relevant eigenvalues of the S-matrix $\tilde{\mathbf{S}}_\mathrm{pp}$ of the FP cavity as a function of the modulation amplitude $M$ at $\omega=\omega_\mathrm{qbic}$. For low $M$, all the eigenvalues lie on the unit circle. As $M$ is increased, a pair of eigenvalues corresponding to the $x$-polarized modes form an exceptional point denoted by $\textrm{EP}_x$. Upon further increasing $M$, one of the eigenvalues approaches $\lambda=0$, marking a CPA point $\textrm{CPA}_x$, while another approaches $\lambda=-\infty$, marking a lasing point $L_x$ (not shown). The $y$-polarized modes also exhibit a similar behavior. The corresponding points for the $y$-polarized modes are denoted by $\textrm{EP}_y$, $\textrm{CPA}_y$, and $\textrm{L}_y$. (d) The absolute value of the eigenvalues shown in (c) as a function of $M$. Here, the EP, CPA, and lasing points for both polarizations are clearly visible. Specifically, the EPs appear at $M=4.64\times10^{-3}$, and the CPA and lasing points occur at $M=6\times10^{-3}$. }
\label{fig:FP-BIC}
\end{figure*}

Structural resonances enhance the interaction of light with time-varying matter. Therefore, strong physical effects due to time modulation can be achieved at extremely small modulation amplitudes. We apply our S-matrix approach to study an FP cavity formed by stacking two identical metasurface layers (see Fig.~\ref{fig:FP-BIC}(a)). Here, as well as in the following example, we assume that each metasurface is made from a square lattice of spheres in the $x-y$ plane. The static susceptibility of the spheres is $\chi_{\mathrm{st}} = 10.68$. Each sphere has a radius of $r = 800~\mathrm{nm}$, and the lattice constant is set to $\Lambda = 3.5r$. The cavity parameters are chosen so that, in the absence of time modulation, it is subwavelength. We provide the values of the convergence parameters $l_\mathrm{max}\,,J\,,\textrm{and }G$ for all the simulations shown in this work in the Supplementary Section~\ref{supp-sec:convergence-params}. 

Under static conditions, the cavity supports FP-BICs \cite{alagappan2025fabry,ustimenko2026singular,Semushev2026robustness}. We plot the transmissivity $T_x$ of the cavity for normally incident $x$-polarized PWs as a function of the frequency $\omega$ and distance $d$ between the layers (see Fig.~\ref{fig:FP-BIC}(b)). We observe a series of resonances with vanishing linewidths. These resonances are FP-BICs, which are formed at a frequency where the individual metasurfaces are perfectly reflecting \cite{alagappan2025fabry}. Since the cavity preserves the $\mathrm{C_4}$ symmetry in the $x-y$ plane, its scattering response is polarization-insensitive at normal incidence \cite{amer2025polarization}. Therefore, such FP-BICs are also observed in the transmission spectrum of the $y$-polarized PWs (see Fig.~\ref{fig:T_xy} in the Supplementary Materials). In the following, we slightly detune the distance $d$ away from the genuine BIC condition, resulting in a quasi-bound state in the continuum (qBIC) configuration. We use $d=12.78r$, which corresponds to a qBIC at the frequency $\omega_\mathrm{qbic}=3.266c_0/\Lambda$. We consider qBICs, since genuine BICs with infinite radiative Q-factors cannot be excited in the structure using external fields due to the absence of radiative coupling.

In the time-varying case, we choose the Floquet frequency and the modulation frequency such that the scattered harmonics corresponding to $j\in\{-1,0\}$ become resonant. In particular, we assume $\omega=\omega_\mathrm{qbic}$ and $\Omega=2\omega_\mathrm{qbic}$. This choice ensures that the scattered harmonics $\omega_{-1}=-\omega_\mathrm{qbic}$ and $\omega_{0}=\omega_\mathrm{qbic}$ lie at the qBIC resonance of the cavity. Furthermore, we assume that the permittivity of each metasurface of the FP cavity is $\varepsilon(t)=1+\chi_\mathrm{st}[1+M\mathrm{cos}(\Omega t)]$.

We show that the resulting time-varying FP cavities exhibit scattering anomalies such as EPs, CPA, and lasing at extremely small modulation amplitudes $M$, with polarization-insensitive operation. For that, we plot the eigenvalues $\lambda$ of the S-matrix $\tilde{\mathbf{S}}_\mathrm{pp}$ in Fig.~\ref{fig:FP-BIC}(c). We show the four relevant eigenvalues of $\tilde{\mathbf{S}}_\mathrm{pp}$, since only these participate in forming the scattering anomalies. We observe that as the modulation is turned on, i.e., $M\gtrsim0$, all the eigenvalues lie on a unit circle. As $M$ is increased, a pair of eigenvalues on the left side (on the unit circle) approach each other, eventually coalescing at an EP. This coalescing involves $x$-polarized modes of the cavity; therefore, we denote the EP as $\textrm{EP}_x$. As $M$ is further increased, the eigenvalues leave the unit circle in pairs. While one of the eigenvalues approaches $\lambda=0$, marking a CPA point ($\textrm{CPA}_x$), the other eigenvalue approaches $\lambda=-\infty$, marking a lasing point ($\textrm{L}_x$, not shown). The coalescing and the subsequent pairwise departure of the eigenvalues are a direct consequence of the pseudounitarity of $\tilde{\mathbf{S}}_\mathrm{pp}$ (see Fig.~\ref{fig:FP-pseudo} in the Supplementary Materials). The eigenvalues of the pseudounitarity matrix $\tilde{\mathbf{S}}_\mathrm{pp}$ form inverse-conjugate pairs, i.e., $\{\lambda,\frac{1}{\lambda^*}\}$, explaining their correlated behavior on the unit circle as well as beyond the EP \cite{globosits2025exceptional}. Due to the polarization-insensitivity of the cavity, EP, CPA, and lasing points are also observed for $y$-polarized eigenmodes denoted by $\textrm{EP}_y$, $\textrm{CPA}_y$, and $\textrm{L}_y$, respectively. We emphasize that to observe such points, $M$ remains perturbative, i.e., $M\ll1$.

To illustrate the discussed scattering anomalies more clearly, we show the absolute values of the relevant eigenvalues in Fig.~\ref{fig:FP-BIC}(d). Again, we observe that when the time modulation is turned on, the absolute values of all four eigenvalues are initially unity. As $M$ increases, at a certain $M$, a pair of eigenvalues ($\lambda_x^\mathrm{CPA}$ and $\lambda_x^\mathrm{L}$) corresponding to the $x$-polarized modes depart from the unity condition forming $\mathrm{EP}_x$. The characteristic square-root dependence of the eigenvalues with $M$ near the EP is also visible. As $M$ is further increased, $|\lambda_x^\mathrm{CPA}|$ asymptotically approaches $0$ while $|\lambda_x^\mathrm{L}|$ approaches $\infty$, marking $\textrm{CPA}_x$ and $\textrm{L}_x$ points, respectively. Specifically, the EP appears at $M=4.64\times10^{-3}$, and the CPA and lasing points occur at $M=6\times10^{-3}$. A similar behavior is observed for the $y$-polarized eigenmodes (with the eigenvalues $\lambda_y^\mathrm{CPA}$ and $\lambda_y^\mathrm{L}$) that mark $\textrm{EP}_y$, $\textrm{CPA}_y$, and $\textrm{L}_y$ points. 

\begin{figure*}
\centerline{\includegraphics[width= 0.5\columnwidth,trim=0.1 0.1 0.1 0.1,clip]{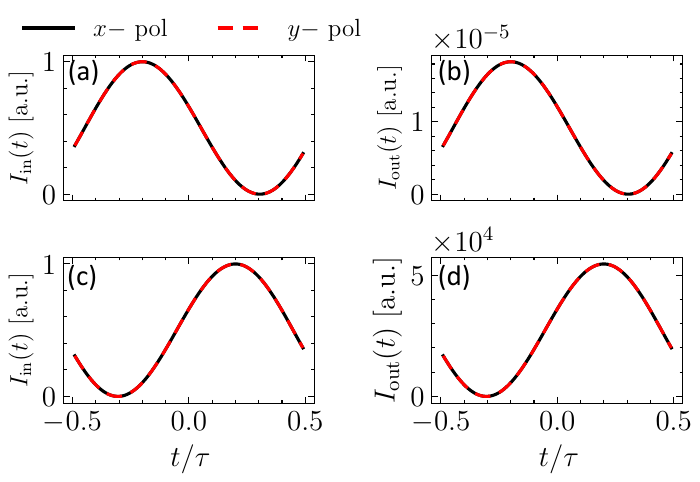}}
\caption{\textbf{Input and output intensities for the CPA and lasing states of the FP cavity.} (a) The space-integrated far-field incident intensity $I_\mathrm{in}(t)$ for the $x$ (solid black curve) and $y$-polarized (dotted red curve) CPA states as a function of time $t$. (b) The corresponding space-integrated far-field outgoing intensity $I_\mathrm{out}(t)$ as a function of time $t$. Here, we observe that the outgoing field is extremely attenuated. (c) and (d) Same as (a) and (b) but for the lasing state. We find that the outgoing field associated with the lasing states is strongly amplified. Here, we evaluate the far-fields at $z=1000\tau c_0$, using $M=6\times10^{-3}$, $\omega=\omega_\mathrm{qbic}$, and $\tau=2\pi/\Omega$.
}
\label{fig:CPA_lasing}
\end{figure*}
We show that the eigenvectors corresponding to the CPA and lasing points are the optimal input states of the system, which, when injected into the cavity, get perfectly absorbed or extremely amplified, respectively. Let us assume that $\tilde{\mathbf{U}}$ is an eigenvector of the matrix $\tilde{\mathbf{S}}_\mathrm{pp}$ with an eigenvalue of $\lambda$. Using $\tilde{\mathbf{U}}$ as the incident state to the scattering system described by $\tilde{\mathbf{S}}_\mathrm{pp}$, the outgoing state is given as $\tilde{\mathbf{U}}_\mathrm{out}=\tilde{\mathbf{S}}_\mathrm{pp} \tilde{\mathbf{U}}=\lambda\tilde{\mathbf{U}}$. Therefore, if $|\lambda|=0$, $\tilde{\mathbf{U}}_\mathrm{out}=0$ signifying CPA, and if $|\lambda|=\infty$, $\tilde{\mathbf{U}}_\mathrm{out}=\infty$ signifying lasing. To demonstrate CPA for the FP cavity, we plot the space-integrated far-field incident intensity $I_\mathrm{in}(t)$ for both $x$- (solid black curve) and $y$-polarized (dotted red curve) CPA states as a function of time $t$ in Fig.~\ref{fig:CPA_lasing}(a). These intensities correspond to the points $\mathrm{CPA}_x$ and $\mathrm{CPA}_y$ with $|\lambda^\mathrm{CPA}_x|=|\lambda^\mathrm{CPA}_y|=0.0043$ at $M=6\times10^{-3}$. The respective far-field outgoing intensities are shown in Fig.~\ref{fig:CPA_lasing}(b). As expected, we observe that $I_\mathrm{out}(t)$ is negligible for all times $t$, indicating that almost all the incident field is absorbed. Similarly, we show the incident intensity for the lasing states in Fig.~\ref{fig:CPA_lasing}(c). These intensities correspond to the points $\mathrm{L}_x$ and $\mathrm{L}_y$ with eigenvalues $|\lambda^\mathrm{L}_x|=|\lambda^\mathrm{L}_y|=233$ at $M=6\times10^{-3}$. The respective far-field outgoing intensities are shown in Fig.~\ref{fig:CPA_lasing}(d). Here, we observe a strong enhancement of the outgoing intensity, as expected.

\subsection*{Nonreciprocal transmission based on symmetry-protected BICs}

\begin{figure*}
\centerline{\includegraphics[width= 1\columnwidth,trim=0.1 0.1 0.1 0.1,clip]{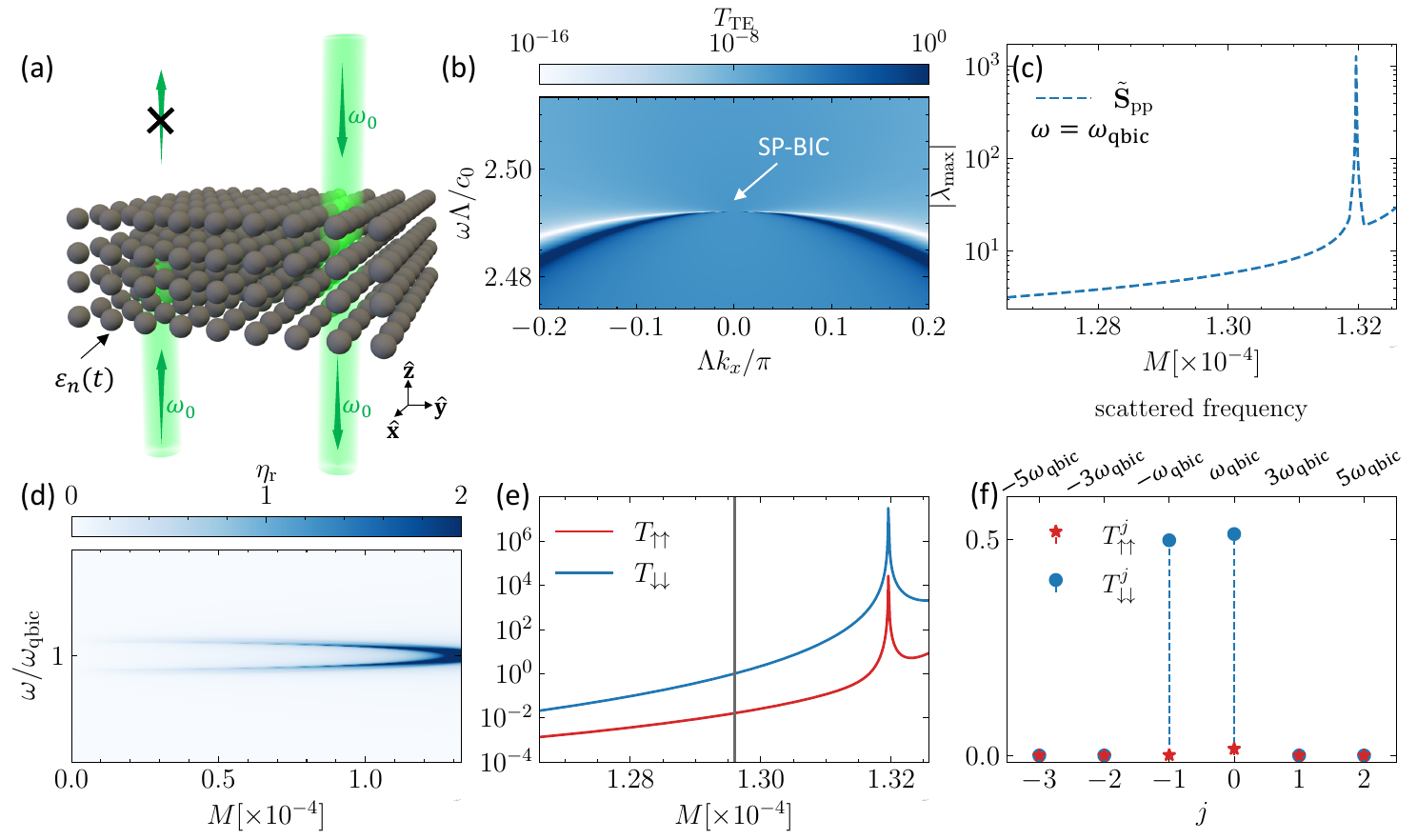}}
\caption{\textbf{Nonreciprocal multilayered metasurface} (a) A multilayered nonreciprocal system formed by stacking four time-varying metasurfaces. The permittivity of each sphere in the $n^\textrm{th}$ layer is $\varepsilon_n(t)=1+\chi_\mathrm{st}[1+M\mathrm{cos}(\Omega t+Kz_n)]$. (b) The transmissivity $T_\mathrm{TE}$ of the multilayered metasurface as a function of frequency $\omega$ and the $x$-component of the in-plane wavevector $k_x$ under static conditions. The metasurface exhibits an SP-BIC at $k_x=0$. (c) The eigenvalue with the maximal absolute value, $\lambda_\mathrm{max}$, of the S-matrix $\tilde{\mathbf{S}}_\mathrm{pp}$ of the metasurface as a function of the modulation amplitude $M$ at $\omega=\omega_\mathrm{qbic}$. The S-matrix has a pole at $M=1.319\times10^{-4}$ signified by a divergent $|\lambda_\mathrm{max}|$. (d) The nonreciprocity parameter $\eta_\mathrm{r}$ as a function of the frequency $\omega$ and modulation amplitude $M$. A non-zero value of $\eta_\mathrm{r}$ implies broken reciprocity. In particular, nonreciprocity is enhanced near $M=1.319\times10^{-4}$ at $\omega=\omega_\mathrm{qbic}$ due to the pole of the S-matrix. (e) The total upward and downward transmissivities, $T_{\uparrow\uparrow}$ and $T_{\downarrow\downarrow}$, for an obliquely incident TE-polarized PW with $k_x=0.0179\pi/\Lambda$ at $\omega=\omega_\mathrm{qbic}$, as a function of $M$. Here, nonreciprocity induces a dissymmetry of two orders of magnitude between $T_{\uparrow\uparrow}$ and $T_{\downarrow\downarrow}$. (f) The harmonic-resolved transmissivity contributions $T^j_{\uparrow\uparrow}$ and $T^j_{\downarrow\downarrow}$ as a function of the harmonic number $j$ and the corresponding scattered frequency for $M=1.296\times10^{-4}$ (marked by the gray line in (e)). Here, the total upward transmissivity $T_{\uparrow\uparrow}=\sum_jT^j_{\uparrow\uparrow}$ is $0.016$, while the total downward transmissivity $T_{\downarrow\downarrow}=\sum_j T^j_{\downarrow\downarrow}$ is $1.01$. Furthermore, the downward transmitted field contains photons of the same energy as the incident fields, i.e., $\hbar\omega_\mathrm{qbic}$. For simplicity, we assume $k_y=0$ in (c)--(f).}
\label{fig:nonreciprocal}
\end{figure*}

We demonstrate that by leveraging structural resonances and engineering the modulation phase $\phi_n$, strong nonreciprocity can be achieved in multilayered metasurfaces at extremely small modulation amplitudes. We consider a scattering structure formed by stacking four identical metasurface layers (see Fig.~\ref{fig:nonreciprocal}(a)). The radius of the spheres is $r=800$~nm, the lattice constant is $\Lambda=2.5r$, and the separation between adjacent layers is $10r$. The resulting multilayered metasurface is subwavelength and supports an SP-BIC under static conditions \cite{koshelev2023bound,Ustimenko2024resonances}. We show the transmissivity $T_\mathrm{TE}$ of the metasurface as a function of frequency $\omega$ and the $x$-component of the in-plane wavevector $k_x$ for TE-polarized incident PWs (see Fig.~\ref{fig:nonreciprocal}(b)). For simplicity, we fix $k_y=0$ for all the simulations presented in this section. We observe a resonance with a vanishing linewidth for $k_x=0$. This resonance is an SP-BIC formed due to the mirror symmetries of the metasurface in the $x-y$ plane \cite{koshelev2018asymmetric}. Next, in the time-varying case, we choose $\Omega=2\omega_\mathrm{qbic}$. Here, the qBIC is attained by slightly detuning from the $k_x=0$ condition. Specifically, we use $k_x=0.0179\pi/\Lambda$, at which the metasurface hosts a qBIC at the frequency $\omega_\mathrm{qbic}=2.49c_0/\Lambda$ (see Fig.~\ref{fig:nonreciprocal}(b)).

A nonreciprocal optical response can be achieved by breaking the time-reversal symmetry in linear systems \cite{asadchy2020tutorial}. In an otherwise homogeneous medium, a traveling-wave permittivity of type $\varepsilon(z,t)=1+\chi_\mathrm{st}[1+M\mathrm{cos}(\Omega t+Kz)]$ is known to exhibit a nonreciprocal response \cite{sounas2017nonreciprocal}. Here, $K$ is the wavenumber of the modulation. To emulate such a permittivity profile for the multilayered metasurface in Fig.~\ref{fig:nonreciprocal}(a), we assume the permittivity of the spheres in the $n^\textrm{th}$ layer to be $\varepsilon_n(t)=1+\chi_\mathrm{st}[1+M\mathrm{cos}(\Omega t+Kz_n)]$, with $z_n$ denoting the spatial coordinate of the $n^\textrm{th}$ layer. Here, $\phi_n=Kz_n$ effectively replaces the continuous spatial phase of a traveling wave with a discrete phase profile. We use $\phi_n=\frac{2.85n\pi}{4}$ with $n\in[0,3]$. The phase profile is chosen so that the nonreciprocity of the resulting structure is significant over the parameter range of interest. Next, we show the eigenvalue with the maximal absolute value, $\lambda_\mathrm{max}$, of the scattering matrix $\tilde{\mathbf{S}}_\mathrm{pp}$ of the multilayered metasurface as a function of $M$ (see Fig.~\ref{fig:nonreciprocal}(c)). Here, we operate at $\omega=\omega_\mathrm{qbic}$, such that $\omega_{-1}=-\omega_\mathrm{qbic}$ and $\omega_{0}=\omega_\mathrm{qbic}$. We observe that $|\lambda_\mathrm{max}|$ diverges at $M=1.319\times10^{-4}$. This divergence indicates a pole of $\tilde{\mathbf{S}}_\mathrm{pp}$. Physically, the pole appears at a modulation amplitude $M$ for which the radiative loss in the qBIC mode is exactly balanced by the time modulation-induced gain \cite{song2026coexistence}. Since, by definition, the qBIC mode has very low radiative loss, a small time modulation-induced gain is needed to compensate for it. Therefore, we observe the divergence of $|\lambda_\mathrm{max}|$ for $M\ll1$. 
The pole of the S-matrix enhances the nonreciprocal response of the metasurface in its vicinity in the parameter space. To quantify nonreciprocity, we plot $\eta_\mathrm{r}=\lVert {\mathbf{V}}\tilde{\mathbf{S}}_\mathrm{pp}^\mathrm{T}{\mathbf{V}}-\tilde{\mathbf{S}}_\mathrm{pp} \rVert_\mathrm{F}$ as a function of the frequency $\omega$ and the modulation amplitude $M$ in Fig.~\ref{fig:nonreciprocal}(d) ($\lVert\,.\rVert_\mathrm{F}$ represents the Frobenius norm). A non-zero value of $\eta_\mathrm{r}$ implies ${\mathbf{V}}\tilde{\mathbf{S}}_\mathrm{pp}^\mathrm{T}{\mathbf{V}}\neq\tilde{\mathbf{S}}_\mathrm{pp}$, and therefore, broken reciprocity. We observe a strong nonreciprocal behavior of the system for near $M=1.319\times10^{-4}$ at $\omega=\omega_\mathrm{qbic}$, as expected. 

Nonreciprocity induces directionality in the system, which leads to an asymmetric transmission response with respect to the incident wave direction. Mathematically, ${\mathbf{V}}\tilde{\mathbf{S}}_\mathrm{pp}^\mathrm{T}{\mathbf{V}}\neq\tilde{\mathbf{S}}_\mathrm{pp}$ implies $|\tilde{\mathbf{S}}^\mathrm{T}_\mathrm{pp,\uparrow\uparrow}|\neq|\tilde{\mathbf{S}}_\mathrm{pp,\downarrow\downarrow}|$ which manifests as direction-dependent transmission (see Eq.~\eqref{eq:Smatrix}). To illustrate such a directional bias, we operate in the parameter range where the metasurface is strongly nonreciprocal. We show the total upward and downward transmissivities, $T_{\uparrow\uparrow}$ and $T_{\downarrow\downarrow}$, for an obliquely incident monochromatic TE-polarized PW with $k_x=0.0179\pi/\Lambda$ at $\omega=\omega_\mathrm{qbic}$, as a function of $M$ in Fig.~\ref{fig:nonreciprocal}(e) (also see Supplementary Section~\ref{supp-sec:RT}). As expected, we observe that $T_{\uparrow\uparrow}$ and $T_{\downarrow\downarrow}$ do not coincide. Furthermore, due to the pole of $\tilde{\mathbf{S}}_\mathrm{pp}$, both transmittances diverge at $M=1.319\times10^{-4}$. In fact, in the vicinity of this divergence, we observe a strong dissymmetry of two orders of magnitude between $T_{\uparrow\uparrow}$ and $T_{\downarrow\downarrow}$. 

To demonstrate that the nonreciprocal device is capable of monochromatic one-way transmission, we choose to operate at $M=1.296\times10^{-4}$ (see the gray vertical line in Fig.~\ref{fig:nonreciprocal}(e)). We plot the harmonic-resolved transmissivity contributions $T^j_{\uparrow\uparrow}$  and $T^j_{\downarrow\downarrow}$ in Fig.~\ref{fig:nonreciprocal}(f) (also see Supplementary Section~\ref{supp-sec:RT}). We observe that when the light is incident from the bottom, the resulting transmissivity contributions $T^j_{\uparrow\uparrow}$ in all the scattered harmonics are negligible. As such, the total upward transmissivity $T_{\uparrow\uparrow}=\sum_jT^j_{\uparrow\uparrow}=0.016$. On the other hand, for the light incident from the top, the field is dominantly transmitted to the frequencies $\pm\omega_\mathrm{qbic}$. In particular, the total downward transmissivity $T_{\downarrow\downarrow}=\sum_jT^j_{\downarrow\downarrow}=1.01$. Therefore, the proposed system exhibits a nonreciprocity-driven one-way transmission response. Specifically, the downward transmitted field only contains photons of the same energy as the incident fields, i.e., $\hbar\omega_\mathrm{qbic}$ (where $\hbar$ is the reduced Planck's constant). Hence, such a nonreciprocal system is suitable for monochromatic operation despite time modulation.

\section*{Discussion}
We have presented the first application of BICs to multilayered time-varying metasurfaces. Leveraging Fabry-Perot BICs in a metasurface-based cavity, we have demonstrated the emergence of scattering anomalies such as exceptional points (EPs), coherent perfect absorption (CPA), and lasing at extremely small modulation amplitudes. The spatial symmetry of the Fabry-Perot cavity in 2D imparts a polarization-insensitive character to these scattering anomalies. In a second example, by utilizing symmetry-protected BICs and breaking the time-reversal symmetry in multilayered time-varying metasurfaces, we have demonstrated the emergence of strong nonreciprocal behavior. By engineering the system parameters, we find that the multilayered metasurface exhibits one-way light transmission at extremely small modulation amplitudes. In fact, despite temporal modulation, the transmitted field remains monochromatic. 

Our work represents an important step toward combining the physics of nanophotonic resonances with time-varying metamaterials. It introduces a platform for investigating rich physical phenomena arising from the interplay between spatial structuring and temporal modulation. We expect our results to be a valuable addition to the field of resonant time-varying nanophotonics.

\backmatter

\bmhead{Supplementary information}
See the supplementary material for more information
\bmhead{Acknowledgements}
The authors would like to thank Nikita Ustimenko and David Globosits for fruitful discussions. They also thank Nikita Ustimenko for proofreading the first draft of the manuscript. P.G. and C.R. are part of the Max Planck School of Photonics, supported by the Bundesministerium für Bildung und Forschung, the Max Planck Society, and the Fraunhofer Society. P.G. and C.R. acknowledge support by the German Research Foundation within the SFB 1173 (project ID no. 258734477). P.G. acknowledges support from the Karlsruhe School of Optics and Photonics (KSOP).

\bmhead{Data availability}
The data supporting the results reported in this paper are available from the authors upon reasonable request.

\bmhead{Authors' Contributions}
P.G. conceived the idea of using bound states in the continuum in multilayered time-varying metasurfaces. P.G. performed all the  theoretical calculations and numerical simulations. M.P. and C.R. supervised the work. P.G. wrote the first draft of the manuscript. All the authors contributed to the discussions of the results and the manuscript preparation.

\bmhead{Competing interests}
The authors declare no competing interests.

\section*{Supplementary Materials}
\setcounter{section}{0}
\renewcommand{\thesection}{S\arabic{section}}
\setcounter{figure}{0}
\renewcommand{\thefigure}{S\arabic{figure}}
\setcounter{equation}{0}
\renewcommand{\theequation}{S\arabic{equation}}

\makeatletter
\renewcommand{\theHsection}{S\arabic{section}}
\renewcommand{\theHequation}{S\arabic{section}.\arabic{equation}}
\renewcommand{\theHfigure}{S\arabic{section}.\arabic{figure}}
\makeatother

\section{Symmetry relation for VSWs}\label{supp-sec:VSW_symmetry}
The following symmetry relation is used to compute various quantities involving negative frequencies \cite{Kim2004symmetry}
\begin{equation}
\label{eq:symmetry_VSW}    
    [\mathbf{F}^{(\iota)}_{lms}(-\kappa^*\mathbf{r})]^*=(-1)^{l+m+\delta_{s 1}} \mathbf{F}^{(\iota)}_{l(-m)s}
    (\kappa\mathbf{r})\,,
\end{equation} 
where $\kappa\in\mathbb{C}$.
\section{Symmetry relation for translation coefficients}\label{supp-sec:C_symmetry}
In this section, we derive the symmetry relation given by Eq.~\eqref{eq:cmatsymm} in the main text. The translation coefficients $c^{\{l m s\},\{ l' m' s'\}}$ for VSWs satisfy \cite{beutel2021efficient,cruzan1962translational,stein1961addition}

\begin{equation}
\label{eq:translation_coeff}
\mathbf{F}^{(3)}_{l' m' s'}(k(\mathbf{r}-\mathbf{R})) 
= \sum_{l m s} 
c^{\{l m s\},\{ l' m' s'\}}(-k\mathbf{R})
\; \mathbf{F}^{(1)}_{l m s}(k\mathbf{r}), \hspace{6pt} \textrm{for }r<R\,.
\end{equation}
Substituting Eq.~\eqref{eq:symmetry_VSW} in Eq.~\eqref{eq:translation_coeff}, we get

\begin{gather*}
\mathbf{F}^{(3)}_{l' m' s'}(k(\mathbf{r}-\mathbf{R})) 
= \sum_{l m s}
c^{\{l m s\},\{ l' m' s'\}}(-k\mathbf{R})
(-1)^{l+m+\delta_{s1}}
\mathbf{F}^{(1)*}_{l (-m) s}(-k^*\mathbf{r})\,,
\intertext{complex conjugating both sides, and substituting, $m \rightarrow -m$ on the right-hand side (R.H.S), we get}\qquad\nonumber
\mathbf{F}^{(3)*}_{l' m' s'}(k(\mathbf{r}-\mathbf{R}))
= \sum_{l m s}
\left[c^{\{l (-m) s\},\{ l' m' s'\}}(-k\mathbf{R})\right]^*
(-1)^{l-m+\delta_{s1}}
\mathbf{F}^{(1)}_{l m s}(-k^*\mathbf{r})\,,
\intertext{using Eq.~\eqref{eq:symmetry_VSW} on the left-hand side (L.H.S), we get}
(-1)^{l'+m'+\delta_{s'1}}
\mathbf{F}^{(3)}_{l'(-m')s'}(-k^*(\mathbf{r}-\mathbf{R}))
=
\sum_{l m s}
\left[c^{\{l (-m) s\},\{ l' m' s'\}}(-k\mathbf{R})\right]^*
(-1)^{l-m+\delta_{s1}}
\mathbf{F}^{(1)}_{l m s}(-k^*\mathbf{r})\,,
\intertext{using Eq.~\eqref{eq:translation_coeff} on the L.H.S, we get}
(-1)^{l'+m'+\delta_{s'1}}
\sum_{l m s}\
c^{\{lms\},\{l' (-m') s'\}}(k^*\mathbf{R})
\mathbf{F}^{(1)}_{l m s}(-k^*\mathbf{r})
=
\sum_{l m s}
\left[c^{\{l (-m) s\},\{ l' m' s'\}}(-k\mathbf{R})\right]^*
(-1)^{l-m+\delta_{s1}}
\mathbf{F}^{(1)}_{l m s}(-k^*\mathbf{r})\,,
\intertext{comparing both sides, and using the linear independence of the VSWs, we get}
(-1)^{l'+m'+\delta_{s'1}}
c^{\{lms\},\{l' (-m') s'\}}(k^*\mathbf{R})
=
(-1)^{l-m+\delta_{s1}}\left[c^{\{l (-m) s\},\{ l' m' s'\}}(-k\mathbf{R})\right]^*\,,
\intertext{therefore,}
c^{\{l m s\},\{ l' (-m') s'\}}(k^*\mathbf{R})
=
(-1)^{l-l'-m-m'+\delta_{s1}-\delta_{s'1}}\left[c^{\{l (-m) s\},\{ l' m' s'\}}(-k\mathbf{R})\right]^*\,,
\intertext{substituting, $m' \rightarrow -m'$, we obtain \nonumber}
c^{\{l m s\},\{ l' m' s'\}}(k^*\mathbf{R})
=
(-1)^{l-l'-m+m'+\delta_{s1}-\delta_{s'1}}\left[c^{\{l(-m)s\},\{l' (-m') s'\}}(-k\mathbf{R})\right]^*\,,
\intertext{substituting, $k \rightarrow k^*$, we get \nonumber}
c^{\{l m s\},\{ l' m' s'\}}(k\mathbf{R})
=
(-1)^{l-l'-m+m'+\delta_{s1}-\delta_{s'1}}\left[c^{\{l(-m)s\},\{l' (-m') s'\}}(-k^*\mathbf{R})\right]^*\,,
\intertext{substituting, $k \rightarrow -k$, we can write\nonumber}
c^{\{l m s\},\{ l' m' s'\}}(-k\mathbf{R})
=
(-1)^{l-l'+m'-m+\delta_{s1}-\delta_{s'1}}\left[c^{\{l(-m)s\},\{l' (-m') s'\}}(-(-k^*)\mathbf{R})\right]^*\,.
\end{gather*}
The final expression is identical to Eq.~\eqref{eq:cmatsymm} in the main text. 
\section{Definition of unit vectors for plane waves}\label{supp-sec:unit_vectors}
For a plane wave of the form $e^{i(\mathbf{k}\cdot\mathbf{r}-\omega t)}\hat{\mathbf{e}}_\alpha(\mathbf{k})$, assuming $(k_x,k_y)\in\mathbb{R}^2$ and $(k_x,k_y)\neq(0,0)$, we use the following definition of the unit vectors \cite{beutel2021efficient}:
\begin{subequations}
\begin{align}
    \hat{\mathbf{e}}_\mathrm{0}(\mathbf{k})&=-i\hat{\bm{\phi}}_\mathbf{k}=i\frac{k_y\hat{\mathbf{x}}-k_x\hat{\mathbf{y}}}{\sqrt{k_x^2+k_y^2}}\,,\\
    \hat{\mathbf{e}}_\mathrm{1}(\mathbf{k})&=i\hat{\mathbf{k}}\times\hat{\mathbf{e}}_\mathrm{0}(\mathbf{k})=-\hat{\bm{\theta}}_\mathbf{k}=\frac{-k_xk_z\hat{\mathbf{x}}-k_yk_z\hat{\mathbf{y}}+(k_x^2+k_y^2)\,\hat{\mathbf{z}}}{k\sqrt{k_x^2+k_y^2}}\,.
    \end{align}
    \end{subequations}
Here, $k=\omega/c_0$. Note that $k<0$ for $\omega<0$. Furthermore, $k_z$ can be complex-valued. When $(k_x,k_y)=(0,0)$, we adopt the convention
\begin{subequations}
\begin{align}
    \hat{\mathbf{e}}_\mathrm{0}(\mathbf{k})&=-i\hat{\mathbf{y}}\,,\\
    \hat{\mathbf{e}}_\mathrm{1}(\mathbf{k})&=\mp\hat{\mathbf{x}}\,, \textrm{for }\Re{(k_z)}\gtrless0\,.
    \end{align}
    \end{subequations}
    
\section{Incident field coefficients in the VSW basis}
We expand a unit PW in a basis of regular VSWs as \cite{beutel2021efficient,beutel2024treams}
\begin{equation}\label{eq:Einc_PW}
e^{i(\mathbf{k}_{j\mathbf{g}\alpha d}\cdot\mathbf{r}-\omega_j t)}\hat{\mathbf{e}}_\alpha(\mathbf{k}_{j\mathbf{g}\alpha d})=\sum_{lm s}{A}^{\mathrm{inc}}_{jlm s}(\omega_j,\mathbf{k}_{j\mathbf{g}\alpha d},\hat{\mathbf{e}}_\alpha(\mathbf{k}_{j\mathbf{g}\alpha d}))\mathbf{F}^{(1)}_{lms}(k_j\mathbf{r})\mathrm{e}^{-i\omega_j t}\,.
\end{equation}
The above expression holds for both $\omega_j>0$ and $\omega_j<0$. However, we can also derive a symmetry relation to compute the coefficients ${A}^{\mathrm{inc}}_{jlm s}$ for $\omega_j<0$ using those for $\omega_j>0$. Using Eq.~\eqref{eq:symmetry_VSW} in Eq.~\eqref{eq:Einc_PW}, we write
\begin{gather}
e^{i(\mathbf{k}_{j\mathbf{g}\alpha d}\cdot\mathbf{r}-\omega_j t)}
\hat{\mathbf{e}}_\alpha(\mathbf{k}_{j\mathbf{g}\alpha d})
=
\sum_{lms}
A^{\mathrm{inc}}_{jlms}(\omega_j,\mathbf{k}_{j\mathbf{g}\alpha d},
\hat{\mathbf{e}}_\alpha(\mathbf{k}_{j\mathbf{g}\alpha d}))
(-1)^{l+m+\delta_{s1}}
\mathbf{F}^{(1)*}_{l(-m)s}(-k_j^*\mathbf{r})
e^{-i\omega_j t},
\\
\intertext{complex conjugating both sides, we obtain}
e^{-i(\mathbf{k}^*_{j\mathbf{g}\alpha d}\cdot\mathbf{r}-\omega^*_j t)}
\hat{\mathbf{e}}_\alpha^*(\mathbf{k}_{j\mathbf{g}\alpha d})
=
\sum_{lms}
A^{\mathrm{inc}*}_{jlms}(\omega_j,\mathbf{k}_{j\mathbf{g}\alpha d},
\hat{\mathbf{e}}_\alpha(\mathbf{k}_{j\mathbf{g}\alpha d}))
(-1)^{l+m+\delta_{s1}}
\mathbf{F}^{(1)}_{l(-m)s}(-k_j^*\mathbf{r})
e^{i\omega^*_j t},
\\
\intertext{expanding the PW on the L.H.S. according to Eq.~\eqref{eq:E_inc} and substituting $m \rightarrow -m$ on the R.H.S, we get}
\sum_{lms}
A^{\mathrm{inc}}_{jlms}(-\omega^*_j,-\mathbf{k}^*_{j\mathbf{g}\alpha d},
\hat{\mathbf{e}}^*_\alpha(\mathbf{k}_{j\mathbf{g}\alpha d}))
\mathbf{F}^{(1)}_{lms}(-k_j^*\mathbf{r})
e^{i\omega^*_j t}
=\nonumber\\
\sum_{lms}
A^{\mathrm{inc}*}_{jl(-m)s}(\omega_j,\mathbf{k}_{j\mathbf{g}\alpha d},
\hat{\mathbf{e}}_\alpha(\mathbf{k}_{j\mathbf{g}\alpha d}))
(-1)^{l-m+\delta_{s1}}
\mathbf{F}^{(1)}_{lms}(-k_j^*\mathbf{r})
e^{i\omega^*_j t},
\\
\intertext{comparing both sides and using the linear independence of VSWs, we find}
A^{\mathrm{inc}}_{jlms}(-\omega^*_j,-\mathbf{k}^*_{j\mathbf{g}\alpha d},
\hat{\mathbf{e}}^*_\alpha(\mathbf{k}_{j\mathbf{g}\alpha d}))=(-1)^{l-m+\delta_{s1}}A^{\mathrm{inc}*}_{jl(-m)s}(\omega_j,\mathbf{k}_{j\mathbf{g}\alpha d},
\hat{\mathbf{e}}_\alpha(\mathbf{k}_{j\mathbf{g}\alpha d}))\,,\\
\intertext{we can rewrite,}
A^{\mathrm{inc}}_{jlms}(\omega_j,-\mathbf{k}^*_{j\mathbf{g}\alpha d},
\hat{\mathbf{e}}^*_\alpha(\mathbf{k}_{j\mathbf{g}\alpha d}))=(-1)^{l-m+\delta_{s1}}A^{\mathrm{inc}*}_{jl(-m)s}(-\omega^*_j,\mathbf{k}_{j\mathbf{g}\alpha d},
\hat{\mathbf{e}}_\alpha(\mathbf{k}_{j\mathbf{g}\alpha d}))\,,
\intertext{assuming $\omega_j\in\mathbb{R}$, we get}
A^{\mathrm{inc}}_{jlms}(\omega_j,-\mathbf{k}^*_{j\mathbf{g}\alpha d},
\hat{\mathbf{e}}^*_\alpha(\mathbf{k}_{j\mathbf{g}\alpha d}))=(-1)^{l-m+\delta_{s1}}A^{\mathrm{inc}*}_{jl(-m)s}(-\omega_j,\mathbf{k}_{j\mathbf{g}\alpha d},
\hat{\mathbf{e}}_\alpha(\mathbf{k}_{j\mathbf{g}\alpha d}))\,
\label{eq:Einc_symmetry}.
\end{gather}
Therefore, the incident field coefficients of the PW $e^{-i(\mathbf{k}^*_{j\mathbf{g}\alpha d}\cdot\mathbf{r}+\omega_j t)}
\hat{\mathbf{e}}^*_\alpha(\mathbf{k}_{j\mathbf{g}\alpha d})$ are linked to that of the PW $e^{i(\mathbf{k}_{j\mathbf{g}\alpha d}\cdot\mathbf{r}-(-\omega_j) t)}\hat{\mathbf{e}}_\alpha(\mathbf{k}_{j\mathbf{g}\alpha d})$ by the symmetry relation in Eq.~\eqref{eq:Einc_symmetry}. We use this relation for computing the incident field coefficients for the negative frequencies, i.e., for $\omega_j<0$ from the corresponding expressions at $\omega_j>0$.

\section{Scattered and outgoing field coefficients in the PW basis}\label{supp-sec:scattered_PW}
The coefficient vectors $\mathbf{A}^\mathrm{inc}$ and $\mathbf{A}^\mathrm{sca}$ of a time-varying metasurface in a VSW basis are linked by the effective T-matrix $\mathbf{T}_\mathrm{eff}(\omega,\mathbf{k}_\parallel)$ as \cite{garg2022modeling}
\begin{eqnarray}
\mathbf{A}^\mathrm{sca} = \mathbf{T}_\mathrm{eff}(\omega,\mathbf{k}_\parallel)\,\mathbf{A}^\mathrm{inc}\, .\label{eq:Tmat}
\end{eqnarray}
Using $\mathbf{A}^\mathrm{sca}$, the scattered field can be calculated as \cite{beutel2021efficient}

\begin{equation}
    \label{eq:MS_sca_VSW}
    \mathbf{E}^\mathrm{sca}(\mathbf{r},t)=\sum_j\left(\sum_{\mathbf{R}}\sum_{lms} A^\mathrm{sca}_{jlms}\mathbf{F}^{(3)}_{lms}(k_j(\mathbf{r}-\mathbf{R}))e^{i\mathbf{k}_\parallel\cdot\mathbf{R}}\right)e^{-i\omega_jt},
\end{equation}
where the sum over $\mathbf{R}$ is needed to take into account the contribution of each sphere in the lattice to the total scattered field. Furthermore, to compute the sum in the parentheses, we use the integral representation of VSWs. For $\omega_j>0$, such a representation is given by \cite{beutel2021efficient} 
\begin{equation}\label{eq:int_VSW}
\begin{aligned}
\mathbf{F}^{(3)}_{lm0}(k_j\mathbf{r}) &= \frac{1}{2\pi i^l} \int \frac{d^2 \mathbf{k}_{\parallel}}{k_j \Gamma_{\mathbf{k}_{\parallel}}}
\, \mathbf{X}_{lm}(\hat{\mathbf{k}}) e^{i \mathbf{k} \cdot \mathbf{r}} , \\
\mathbf{F}^{(3)}_{lm1}(k_j\mathbf{r}) &= \frac{1}{2\pi i^{\,l-1}} \int \frac{d^2 \mathbf{k}_{\parallel}}{k_j \Gamma_{\mathbf{k}_{\parallel}}}
\, \hat{\mathbf{k}} \times \mathbf{X}_{lm}(\hat{\mathbf{k}}) e^{i \mathbf{k} \cdot \mathbf{r}}\,, \\
\text{with } \mathbf{k} &= \mathbf{k}_{\parallel} \pm \hat{\mathbf{z}} \Gamma_{\mathbf{k}_{\parallel}}\,, 
\quad \text{for } z \gtrless 0,\,
\text{where } \Gamma_{\mathbf{k}_{\parallel}}=\sqrt{k^2_j-\mathbf{k}^2_\parallel}\,.
\end{aligned}
\end{equation}
For $\omega_j<0$ (assuming $\omega_j\in\mathbb{R}$), we use the symmetry relation given by Eq.~\eqref{eq:symmetry_VSW} in Eq.~\eqref{eq:int_VSW}. Therefore, in this case, the integral representation is given by 

\begin{equation}\label{eq:int_VSW_neg}
\begin{aligned}
\mathbf{F}^{(3)}_{lm0}(k_j \mathbf{r}) 
&= \frac{(-1)^{l+m+\delta_{01}}}{2\pi (-i)^l} 
\int \frac{d^2 \mathbf{k}_{\parallel}}{(-k_j) \, \Gamma_{\mathbf{k}_{\parallel}}^*}
\, \mathbf{X}_{l(-m)}^*(\hat{\mathbf{k}})\, e^{-i \mathbf{k}^* \cdot \mathbf{r}}\,, \\
\mathbf{F}^{(3)}_{lm1}(k_j \mathbf{r}) 
&= \frac{(-1)^{l+m+\delta_{11}}}{2\pi (-i)^{\,l-1}} 
\int \frac{d^2 \mathbf{k}_{\parallel}}{(-k_j) \, \Gamma^*_{\mathbf{k}_{\parallel}}}
\, \hat{\mathbf{k}}^* \times \mathbf{X}^*_{l(-m)}(\hat{\mathbf{k}})\, e^{-i \mathbf{k}^* \cdot \mathbf{r}}\,, \\
\text{with } \mathbf{k} 
&= \mathbf{k}_{\parallel} \pm \hat{\mathbf{z}}\, \Gamma_{\mathbf{k}_{\parallel}}, 
\quad \text{for } z \gtrless 0,\,
\text{where } \Gamma_{\mathbf{k}_{\parallel}}=\sqrt{(-k_j)^2-\mathbf{k}^2_\parallel}\,.
\end{aligned}
\end{equation}

Substituting Eqs.~\eqref{eq:int_VSW} in Eq.~\eqref{eq:MS_sca_VSW} and simplifying the integrals using Poisson's sum formula, the scattered field for $\omega_j>0$ is written as
\begin{equation}
\label{eq:E_sca_pos}
\mathbf{E}^\mathrm{sca}(\mathbf{r},t)
= \sum_{\substack{j\mathbf{g}\alpha d\\ \textrm{s.t. }\omega_j>0}} 
U^\mathrm{sca}_{j\mathbf{g}\alpha d}\,
e^{i(\mathbf{k}_{j\mathbf{g}\alpha d} \cdot \mathbf{r}-\omega_jt)}\,\hat{\mathbf{e}}_{\alpha}(\mathbf{k}_{j\mathbf{g}\alpha d})\,,
\end{equation}
Here,
\begin{subequations}
\begin{align}
U^\mathrm{sca}_{j\mathbf{g}\alpha d}&=\sum_{lms}\frac{2\pi \gamma_{lm} e^{i m \phi_{\mathbf{k}_{j\mathbf{g}\alpha d}}}}
{k_j |A_\mathrm{u}|\,i^{\,l+1} \Gamma_{\mathbf{k}_{j\mathbf{g}\alpha d}}}
a_{jlms}^{\mathbf{k}_{j\mathbf{g}\alpha d}}\,,\\
\mathbf{k}_{j\mathbf{g}\alpha d} &= (\mathbf{k}_\parallel+\mathbf{g})+\hat{\mathbf{z}}(-1)^{\delta_{d\downarrow}}\Gamma_{\mathbf{k}_{j\mathbf{g}\alpha d}}\,, \\
\Gamma_{\mathbf{k}_{j\mathbf{g}\alpha d}} &= \sqrt{k_j^2-(\mathbf{k}_\parallel+\mathbf{g})^2}\,, \\
a_{jlm0}^{\mathbf{k}_{j\mathbf{g}\alpha d}} &= A^\mathrm{sca}_{jlm0}\frac{\partial}{\partial \theta}P^m_l(\cos\theta)\bigg|_{\theta=\theta_{\mathbf{k}_{j\mathbf{g}\alpha d}}}
+ A^\mathrm{sca}_{jlm1}\frac{m}{\sin\theta}P^m_l(\cos\theta)\bigg|_{\theta=\theta_{\mathbf{k}_{j\mathbf{g}\alpha d}}}\,, \\
a_{jlm1}^{\mathbf{k}_{j\mathbf{g}\alpha d}} &= A^\mathrm{sca}_{jlm0}\frac{m}{\sin\theta}P^m_l(\cos\theta)\bigg|_{\theta=\theta_{\mathbf{k}_{j\mathbf{g}\alpha d}}}
+ A^\mathrm{sca}_{jlm1}\frac{\partial}{\partial \theta}P^m_l(\cos\theta)\bigg|_{\theta=\theta_{\mathbf{k}_{j\mathbf{g}\alpha d}}}\,, \\
\gamma_{lm} &= i\sqrt{\frac{(l-m)!}{(l+m)!}\frac{(2l+1)}{4\pi l(l+1)}}\,,\\
\theta_{{\mathbf{k}}_{j\mathbf{g}\alpha d}}&=\mathrm{arccos}\left({\frac{k_{z;j\mathbf{g}\alpha d}}{k_j}}\right)\,,\\
\phi_{{\mathbf{k}}_{j\mathbf{g}\alpha d}}&=\mathrm{arctan}2\left(\frac{k_{y;j\mathbf{g}\alpha d}}{k_{x;j\mathbf{g}\alpha d}}\right)\,,
\end{align}
\end{subequations}
\noindent where $|A_\mathrm{u}|$ is the area of the unit cell of the metasurface lattice.

Similarly, using Eqs.~\eqref{eq:int_VSW_neg} in Eq.~\eqref{eq:MS_sca_VSW}, the scattered field for $\omega_j<0$ is written as
\begin{equation}
\label{eq:E_sca_neg}
\mathbf{E}^\mathrm{sca}(\mathbf{r},t)
= \sum_{\substack{j\mathbf{g}\alpha d\\ \textrm{s.t. }\omega_j<0}}
U^\mathrm{sca}_{j\mathbf{g}\alpha d}\,
e^{-i(\mathbf{k}^{\,*}_{j\mathbf{g}\alpha d} \cdot \mathbf{r}+\omega_j t)}\,\hat{\mathbf{e}}^*_{\alpha}(\mathbf{k}_{j\mathbf{g}\alpha d})\,,
\end{equation}
where
\begin{subequations}
\label{eS_kneg}
\begin{align}
U^\mathrm{sca}_{j\mathbf{g}\alpha d}&= \sum_{lms}{\frac{(-1)^{l+m} 2\pi \gamma^*_{l(-m)}
e^{i m \phi^*_{\mathbf{k}_{j\mathbf{g}\alpha d}}}}
{(-k_j) |A_\mathrm{u}|\, (-i)^{\,l+1} \Gamma^*_{\mathbf{k}_{j\mathbf{g}\alpha d}}}
a_{jlms}^{\mathbf{k}_{j\mathbf{g}\alpha d}}}\,,\\
\mathbf{k}_{j\mathbf{g}\alpha d} &= (-\mathbf{k}_\parallel+\mathbf{g})+\hat{\mathbf{z}}(-1)^{\delta_{d\downarrow}}\Gamma_{\mathbf{k}_{j\mathbf{g}\alpha d}}\,, \\
\Gamma_{\mathbf{k}_{j\mathbf{g}\alpha d}} &= \sqrt{(-k_j)^2-(-\mathbf{k}_\parallel+\mathbf{g})^2}\,, \\
a_{jlm0}^{\mathbf{k}_{j\mathbf{g}\alpha d}} &= A^\mathrm{sca}_{jlm0}\left[\frac{\partial}{\partial \theta}P^{-m}_l(\cos\theta)\bigg|_{\theta=\theta_{\mathbf{k}_{j\mathbf{g}\alpha d}}}\right]^* 
- A^\mathrm{sca}_{jlm1}\left[\frac{m}{\sin\theta}P^{-m}_l(\cos\theta)\bigg|_{\theta=\theta_{\mathbf{k}_{j\mathbf{g}\alpha d}}}\right]^*\,, \\
a_{jlm1}^{\mathbf{k}_{j\mathbf{g}\alpha d}} &= A^\mathrm{sca}_{jlm0}\left[\frac{m}{\sin\theta}P^{-m}_l(\cos\theta)\bigg|_{\theta=\theta_{\mathbf{k}_{j\mathbf{g}\alpha d}}}\right]^* 
- A^\mathrm{sca}_{jlm1}\left[\frac{\partial}{\partial \theta}P^{-m}_l(\cos\theta)\bigg|_{\theta=\theta_{\mathbf{k}_{j\mathbf{g}\alpha d}}}\right]^*\,,\\
\theta_{{\mathbf{k}}_{j\mathbf{g}\alpha d}}&=\mathrm{arccos}\left({\frac{k_{z;j\mathbf{g}\alpha d}}{-k_j}}\right)\,,\\
\phi_{{\mathbf{k}}_{j\mathbf{g}\alpha d}}&=\mathrm{arctan}2\left(\frac{k_{y;j\mathbf{g}\alpha d}}{k_{x;j\mathbf{g}\alpha d}}\right)\,.
\end{align}
\end{subequations}
Note that for $\omega_j>0$ as well as $\omega_j<0$, the scattered field propagates upwards (i.e., in $+z$ direction) for $z>0$, and it propagates downwards (i.e., in $-z$ direction) for $z<0$. Finally, the output field of the metasurface can be written as \cite{beutel2021efficient}
\begin{equation}
    \mathbf{E}^\mathrm{out}(\mathbf{r},t)=\mathbf{E}^\mathrm{sca}(\mathbf{r},t)+\mathbf{E}^\mathrm{inc}(\mathbf{r},t)\,.
\end{equation}
In particular, the output field coefficients $U^\mathrm{out}_{j\mathbf{g}\alpha d}$ defined in Eq.~\eqref{eq:Smat_fields} are given by 
\begin{equation}
    U^\mathrm{out}_{j\mathbf{g}\alpha d}=U^\mathrm{sca}_{j\mathbf{g}\alpha d}+U^\mathrm{inc}_{j\mathbf{g}\alpha d}\,.
\end{equation}
Finally, we note that computing the S-matrix $\mathbf{S}(\omega,\mathbf{k}_\parallel)$ from the effective T-matrix $\mathbf{T}_\mathrm{eff}(\omega,\mathbf{k}_\parallel)$ requires evaluating the coefficients $\mathbf{U}^\mathrm{out}$ corresponding to incident unit PWs (see Eq.~\eqref{eq:Smat_fields}). The $\mathbf{U}^\mathrm{out}$ vector for each incident PW forms a column of $\mathbf{S}(\omega,\mathbf{k}_\parallel)$. We show the corresponding algorithm in Fig.~\ref{fig:flowchart}.

\section{S-matrix star product}\label{supp-sec:star_product}
The S-matrix star product for the layers described by the S-matrices $\mathbf{S}^\mathrm{(a)}$ and $\mathbf{S}^\mathrm{(b)}$ is given by \cite{beutel2021efficient,Rumpf2011improved,Redheffer1959inequalities}

\begin{align}
    \mathbf{S}&=\mathbf{S}^\mathrm{(a)}\star\mathbf{S}^\mathrm{(b)},\textrm{where}\nonumber\\
    \mathbf{S}_{\uparrow\uparrow}&= \mathbf{S}_{\uparrow\uparrow}^\mathrm{(b)}(\mathbf{I}-\mathbf{S}_{\uparrow\downarrow}^\mathrm{(a)}\mathbf{S}_{\downarrow\uparrow}^\mathrm{(b)})^{-1}\mathbf{S}_{\uparrow\uparrow}^\mathrm{(a)}\,, \nonumber \\
    \mathbf{S}_{\uparrow\downarrow}&= \mathbf{S}_{\uparrow\downarrow}^\mathrm{(b)}+\mathbf{S}_{\uparrow\uparrow}^\mathrm{(b)}\mathbf{S}_{\uparrow\downarrow}^\mathrm{(a)}(\mathbf{I}-\mathbf{S}_{\downarrow\uparrow}^\mathrm{(b)}\mathbf{S}_{\uparrow\downarrow}^\mathrm{(a)})^{-1}\mathbf{S}_{\downarrow\downarrow}^\mathrm{(b)}\,, \nonumber\\
    \mathbf{S}_{\downarrow\uparrow}&= \mathbf{S}_{\downarrow\uparrow}^\mathrm{(a)}+\mathbf{S}_{\downarrow\downarrow}^\mathrm{(a)}\mathbf{S}_{\downarrow\uparrow}^\mathrm{(b)}(\mathbf{I}-\mathbf{S}_{\uparrow\downarrow}^\mathrm{(a)}\mathbf{S}_{\downarrow\uparrow}^\mathrm{(b)})^{-1}\mathbf{S}_{\uparrow\uparrow}^\mathrm{(a)}\,, \nonumber\\
    \mathbf{S}_{\downarrow\downarrow}&=
    \mathbf{S}_{\downarrow\downarrow}^\mathrm{(a)}(\mathbf{I}-\mathbf{S}_{\downarrow\uparrow}^\mathrm{(b)}\mathbf{S}_{\uparrow\downarrow}^\mathrm{(a)})^{-1}\mathbf{S}_{\downarrow\downarrow}^\mathrm{(b)}\,.\label{eq:couplingS}
\end{align}
Here, the layer described by $\mathbf{S}^\mathrm{(b)}$ is on top of the layer described by $\mathbf{S}^\mathrm{(a)}$. In general, the determinants of the matrices inside the parentheses are non-zero. A vanishing determinant defines the condition for obtaining the eigenmodes of the composite scattering system \cite{globosits2025exceptional}. Furthermore, the S-matrix star product is associative \cite{MISTIRI1986thestar}.  Note that we have omitted the argument $(\omega,\mathbf{k}_\parallel)$ of the S-matrices for brevity.

\section{Reflectivity and transmissivity of time-varying metasurfaces}\label{supp-sec:RT}
The instantaneous Poynting vector for an arbitrary field distribution characterized by complex-valued electric field $\mathbf{E}(\mathbf{r},t)$ and magnetic field strength $\mathbf{H}(\mathbf{r},t)$ is given by 
\begin{align}
     \mathbf{S}(\mathbf{r},t)=\mathbf{E}_\mathrm{r}(\mathbf{r},t)\times\mathbf{H}_\mathrm{r}(\mathbf{r},t)\,,
\end{align}
where $\mathbf{E}_\mathrm{r}(\mathbf{r},t)=\Re\{\mathbf{E}(\mathbf{r},t)\}$ and $\mathbf{H}_\mathrm{r}(\mathbf{r},t)=\Re\{\mathbf{H}(\mathbf{r},t)\}$.
The space and time-averaged power flux passing through a plane of infinite area is computed by integrating the Poynting vector $\mathbf{S}(\mathbf{r},t)$ as
\begin{align}\label{eq:poynting_integration}
     \langle\mathbf{S}\rangle=\lim_{|A|\to\infty}\,\frac{1}{|A|}\frac{1}{\tau_d}\int_{A}\int_{\tau_\mathrm{d}}\mathbf{S}(\mathbf{r},t)\cdot\hat{\mathbf{z}}\,dt\,dA'\,,
\end{align}
where the area $A$ is assumed to be in the $x-y$ plane, and $\tau_\mathrm{d}$ is the integration time of the detector. 

For a monochromatic upward propagating incident PW to the time-varying metasurface, the space and time-averaged upward propagating power flux is given by integrating the corresponding Poynting vector $\mathbf{S}_\mathrm{inc}$. Assuming that the electric field of the incident PW has the form $\mathbf{E}_\mathrm{inc}(\mathbf{r},t)=U^\mathrm{inc}
e^{i(\mathbf{k}\cdot\mathbf{r}-\omega t)}\hat{\mathbf{e}}_\mathrm{inc}(\mathbf{k})$ (see Eq.~\eqref{eq:E_in}), and performing the integration according to Eq.~\eqref{eq:poynting_integration}, we get
\begin{align}
     \langle\mathbf{S}_\mathrm{inc}\rangle= \frac{|U^\mathrm{inc}|^2}{2Z_0|k|}|\Re\{k_{z}\}|\,,
\end{align}
Here, $\mathbf{k}=\mathbf{k}_\parallel+k_z\hat{\mathbf{z}}$ with $k_z>0$, $Z_0$ is the free-space impedance, and the spatial integration is taken in the limit sense described in Eq.~\eqref{eq:poynting_integration}, over an infinite plane located spatially below the metasurface. 

Assuming that the incident PW gives rise to the outgoing field of the form given in Eq.~\eqref{eq:E_out}, the space and time-averaged power fluxes for the corresponding reflected and transmitted fields are given by
\begin{subequations}\label{eq:ref-tra-S}
\begin{align}
\langle\mathbf{S}_\mathrm{ref}\rangle=\frac{1}{2Z_0}\sum_j\sum_{\mathbf{g}\alpha}\frac{1}{|k_j|}|U^\mathrm{out}_{j\mathbf{g}\alpha \downarrow}|^2\,|\Re\{k_{z;j\mathbf{g}\alpha\downarrow}\}|\,,\\
\langle\mathbf{S}_\mathrm{tra}\rangle=\frac{1}{2Z_0}\sum_j\sum_{\mathbf{g}\alpha}\frac{1}{|k_j|}|U^\mathrm{out}_{j\mathbf{g}\alpha \uparrow}|^2\,|\Re\{k_{z;j\mathbf{g}\alpha\uparrow}\}|\,,
\end{align}
\end{subequations}
where the spatial integration for the reflected fields is taken over an infinite plane spatially below, and that for the transmitted fields is taken over an infinite plane spatially above the metasurface. Furthermore, we assume that the in-plane wavevector $\mathbf{k}_\parallel\neq0$ for the incident wave. Such a choice allows us to write the reflected and transmitted power fluxes as an incoherent sum of the fluxes in individual frequencies in Eq.~\eqref{eq:ref-tra-S}. In case $\mathbf{k}_\parallel=0$, the contributions of the terms arising due to the interference between the positive and negative frequencies are non-negligible. Such an interference is discussed in more detail in \cite{hendry2025effects}.

Finally, we can write the total reflectivity (transmissivity) as the ratio of the power fluxes of the reflected (transmitted) and the incident fields.

\begin{subequations}\label{supp-eq:RT}
\begin{align}
\textrm{reflectivity}=\frac{\langle\mathbf{S}_\mathrm{ref}\rangle}{\langle\mathbf{S}_\mathrm{inc}\rangle}=\sum_jR_j=\sum_{j}{\left(\frac{{\sum_{\mathbf{g}\alpha}\frac{1}{|k_j|}|U^\mathrm{out}_{j\mathbf{g}\alpha \downarrow}|^2\,|\Re\{k_{z;j\mathbf{g}\alpha\downarrow}\}|}}{\frac{|U^\mathrm{inc}|^2}{|k|}|\Re\{k_{z}\}|}\right)}\,,\\
\textrm{transmissivity}=\frac{\langle\mathbf{S}_\mathrm{tra}\rangle}{\langle\mathbf{S}_\mathrm{inc}\rangle}=\sum_jT_j=\sum_{j}{\left(\frac{{\sum_{\mathbf{g}\alpha}\frac{1}{|k_j|}|U^\mathrm{out}_{j\mathbf{g}\alpha \uparrow}|^2\,|\Re\{k_{z;j\mathbf{g}\alpha\uparrow}\}|}}{\frac{|U^\mathrm{inc}|^2}{|k|}|\Re\{k_{z}\}|}\right)}\,.
\end{align}
\end{subequations}
The expressions for the reflectivity and transmissivity of a downward-propagating incident PW can be computed in a similar fashion. 
\section{Details of the convergence parameters used during simulations}\label{supp-sec:convergence-params}
For performing the simulations throughout the article, we used various convergence parameters. These parameters are the maximum multipolar order $l_\mathrm{max}$, the integer $J$, and the total number of diffraction orders $G$. Below, we list the numerical values of these parameters.\\

\noindent Figures~2(b): $l_\mathrm{max}=3$, $G=1$ (static case).\\
\noindent Figures~2(c)--(e): $l_\mathrm{max}=5$, $J=2$, $G=9$.\\
\noindent Figure~3: $l_\mathrm{max}=5$, $J=2$, $G=9$.\\
\noindent Figures~4(b): $l_\mathrm{max}=3$, $G=1$ (static case).\\
\noindent Figures~4(c)--(e): $l_\mathrm{max}=10$, $J=2$, $G=5$.\\
\noindent Figures~4(f): $l_\mathrm{max}=10$, $J=3$, $G=5$.\\
\noindent Figure~S1: $l_\mathrm{max}=3$, $G=1$ (static case).\\
\noindent Figure~S2: $l_\mathrm{max}=5$, $J=2$, $G=9$.\\

\begin{figure*}
\centerline{\includegraphics[width= 1\columnwidth,trim=0.1 0.1 0.1 0.1,clip]{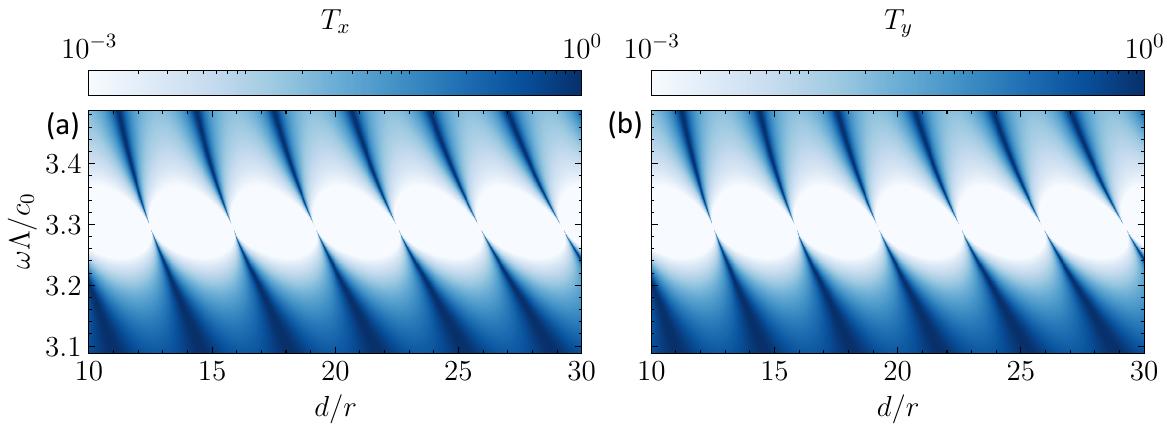}}
\caption{(a)--(b) The transmissivities $T_x$ and $T_y$ of the FP cavity as a function of frequency $\omega$ and the inter-layer distance $d$ under static conditions for the $x$-polarized and $y$-polarized normally incident PWs, respectively. Due to the mirror-symmetry of the FP cavity in the $x-y$ plane, $T_x$ and $T_y$ are identical.}
\label{fig:T_xy}
\end{figure*} 

\begin{figure*}
\centerline{\includegraphics[width= 0.5\columnwidth,trim=0.1 0.1 0.1 0.1,clip]{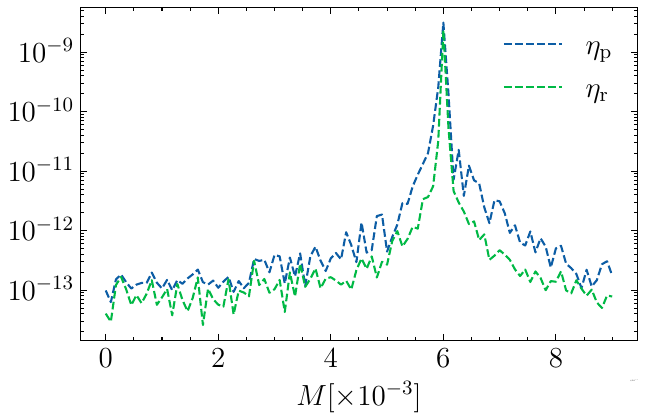}}
\caption{The pseudounitarity and reciprocity parameters $\eta_\mathrm{p}$ and $\eta_\mathrm{r}$ for the FP cavity as a function of the modulation amplitude $M$. We define $\eta_\mathrm{p}=\lVert \tilde{\mathbf{S}}_\mathrm{pp}^\dagger{\mathbf{V}}\tilde{\mathbf{S}}_\mathrm{pp}-{\mathbf{V}} \rVert_\mathrm{F}$, and $\eta_\mathrm{r}=\lVert {\mathbf{V}}\tilde{\mathbf{S}}_\mathrm{pp}^\mathrm{T}{\mathbf{V}}-\tilde{\mathbf{S}}_\mathrm{pp} \rVert_\mathrm{F}$ (where $\lVert\,.\rVert_\mathrm{F}$ represents the Frobenius norm). Here, $\eta_\mathrm{p}\approx0$ and $\eta_\mathrm{r}\approx0$, signifying that the matrix $\tilde{\mathbf{S}}_\mathrm{pp}$ satisfies both the pseudounitarity and reciprocity conditions.}
\label{fig:FP-pseudo}
\end{figure*}

\begin{figure*}
\centerline{\includegraphics[width= 0.25\columnwidth,trim=0.1 0.1 0.1 0.1,clip]{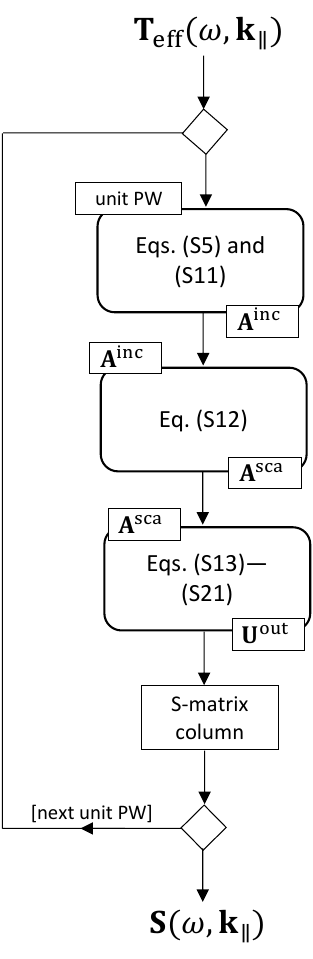}}
\caption{A flow chart illustrating the algorithm to evaluate the S-matrix $\mathbf{S}(\omega,\mathbf{k}_\parallel)$ from the effective T-matrix $\mathbf{T}_\mathrm{eff}(\omega,\mathbf{k}_\parallel)$ of a time-varying metasurface.}
\label{fig:flowchart}
\end{figure*}

\bibliography{references_all}
\end{document}